%% file: 25TWC_neural_positioning.tex
\documentclass[journal]{IEEEtran}

\usepackage{cite}
\usepackage{amsmath,amssymb,amsfonts,amsthm}
\usepackage{graphicx}
\usepackage{textcomp}
\usepackage{hyperref}
\usepackage{booktabs}
\usepackage{multirow}
\usepackage{float}
\usepackage{xcolor}
\usepackage{microtype}
\usepackage{tikz}
\usepackage[caption=false,font=footnotesize]{subfig}
\usepackage{balance}
\usepackage{mdframed}

\theoremstyle{rem}
\newtheorem{rem}{Remark}

\input{macros/vmr-symbols-vecbold}
\input{macros/standard-macros}

\input{macros/defs}

\newcommand{\eg}[1]{\textcolor{green}{\bf [eg: #1]}}

\begin{document}

\title{Neural Positioning Without External Reference}

\author{%
    Till-Yannic M\"uller, Frederik Zumegen, Reinhard Wiesmayr, 
    Emre G\"on\"ulta\c{s}, and Christoph Studer\thanks{This work was supported in part by the Swiss National Science Foundation (SNSF) grant 200021\_207314 and by CHIST-ERA grant for the project CHASER (CHIST-ERA-22-WAI-01) through the SNSF grant 20CH21\_218704. We acknowledge NVIDIA for their sponsorship of this research.}
    \thanks{TM, FZ, RW, and CS are with the Department of Information Technology and Electrical Engineering, ETH Zurich, Switzerland. EG is with Ericsson, Austin, TX 78703, USA.
    (e-mail: tilmueller@ethz.ch, fzumegen@iis.ee.ethz.ch, wiesmayr@iis.ee.ethz.ch,  emre.gonultas@ericsson.com, and studer@ethz.ch)
}
}

\maketitle

\begin{abstract}
Channel state information (CSI)-based user equipment (UE) positioning with neural networks---referred to as neural positioning---is a promising approach for accurate off-device UE localization. Most existing methods train their neural networks with ground-truth position labels obtained from external reference positioning systems, which requires costly hardware and renders label acquisition difficult in large areas. 
In this work, we propose a novel neural positioning pipeline that avoids the need for any external reference positioning system. Our approach trains the positioning network only using CSI acquired off-device and relative displacement commands executed on commercial off-the-shelf (COTS) robot platforms, such as robotic vacuum cleaners---such an approach enables inexpensive training of accurate neural positioning functions over large areas. 
We evaluate our method in three real-world scenarios, ranging from small line-of-sight (LoS) areas to larger non-line-of-sight (NLoS) environments, using CSI measurements acquired in  IEEE 802.11 \WiFi and 5G New Radio (NR) systems. Our experiments demonstrate that the proposed neural positioning pipeline achieves UE localization accuracies close to state-of-the-art methods that require externally acquired high-precision ground-truth position labels for training.
\end{abstract}



\section{Introduction}

\IEEEPARstart{U}{ser} equipment (UE) positioning at the infrastructure access points (APs) or basestations is expected to play a central role in next-generation wireless networks~\cite{6g_localization_sensing}.
A particularly promising approach is UE positioning based on measured channel-state information (CSI), which enables accurate positioning alongside regular data transmission---without utilizing additional resources of the wireless spectrum. 
Accurate UE positioning can, for example, assist (and improve) critical physical layer and network management tasks such as rate adaptation, resource allocation, and handover optimization~\cite{haghrah2023survey}.

UE positioning can be performed either \emph{on-device}, where the UE estimates its own position, or \emph{off-device}, where the position is inferred at the network operator side, e.g., at infrastructure APs, based on signals transmitted by the UE~\cite{huang2024attackDefendOffDevicePos}.
While on-device positioning via global navigation satellite systems (GNSSs) is widespread, GNSSs do not work well indoors or in dense urban scenarios~\cite{Granados2012GNSSIndoor}. 
Furthermore, network operators do, in general, not have access to GNSS position information. 
In contrast, off-device UE positioning with neural networks that process measured CSI acquired at infrastructure APs---referred to as \emph{neural positioning}---enables network operators to acquire accurate UE position information, even indoors and in dense urban scenarios~\cite{Chapre2014CSIFingerprinting,He2016CSIFingerprinting,wang2017indoorLocalDeepLearning,wu2013csiIndoorLocal,Arnold2019CSIFingerprinting,Savic2015Figerprinting,Vieira2017Fingerprinting,Gönültaş2021probabilityFusion}.
Such systems, however, typically require training of the positioning functions (commonly implemented with neural networks) from large CSI datasets labeled with ground-truth location data, which is typically acquired with external and highly-accurate reference positioning systems.
External reference positioning systems are costly and difficult to set up (and to maintain) in large and complex environments.
Furthermore, such systems need to be redeployed whenever the positioning neural networks must be retrained (e.g., because of changes in the environment). 
In addition, the need for external reference positioning systems to train neural positioning functions somewhat defies the purpose of deploying positioning systems that leverage the existing wireless communication infrastructure in the first place.

\subsection{Contributions}
We propose a neural positioning pipeline that enables learning of accurate UE positioning functions without requiring an external reference positioning system.
Our main contributions are summarized as follows:
\begin{itemize}
\item We propose a neural positioning pipeline that trains accurate positioning neural networks solely using externally measured CSI at distributed APs and relative displacement commands provided to commercial off-the-shelf (COTS) robot platforms (e.g., cleaning robots) that pass through the area of interest during the measurement campaign. 
\item We improve positioning accuracy through a novel training procedure, which builds upon a triangle displacement loss extracted from relative displacement commands. 
\item We acquire CSI data and relative displacement commands through real-world measurements with COTS devices and standard wireless communication infrastructure, namely IEEE 802.11 \WiFi and 5G New Radio (NR) systems. 
\item We validate the effectiveness of our approach with experiments in three different scenarios, ranging from small line-of-sight (LoS) areas to larger non-line-of-sight (NLoS) environments. All datasets will be made publicly available after the peer-review process. 
\item We compare the positioning accuracy of our approach to several baselines, including neural positioning that requires external reference positioning systems, and demonstrate that our proposed method achieves comparable accuracy. 
\end{itemize}

\subsection{Relevant Prior Art}

Off-device positioning based on wireless communication infrastructure has been extensively studied within the last decade~\cite{yang2013fromRSSItoCSI}. Traditional methods rely (i) on received signal strength indicator (RSSI) measurements~\cite{Zou2017RSSI,Yang2009RSSI}, which estimate distance based on signal attenuation, or (ii) on time- and angle-based measurements, such as time-of-arrival (ToA), time-difference-of-arrival (TDoA), and angle-of-arrival (AoA)~\cite{Qi2006ToARSSI,Chen2020AoA,ahadi2025displacementCC}, where the UEs are positioned through trilateration or triangulation. 
Such approaches are at the mercy of multipath propagation because LoS components are difficult to identify or simply may not be present. 
In contrast, our proposed method works robustly and accurately even under complex multipath channel conditions.  

Alternative off-device positioning methods rely on CSI measured at infrastructure APs (or basestations), which enables statistical-deterministic approaches, such as modeling the probabilistic relationship between CSI and UE locations using particle filtering~\cite{Shen2020ParticleFilterBluetooth,deutschmann2024directPosParticleFilter}. 
Such approaches require accurate modeling, as they heavily rely on assumptions about motion and signal distributions. 
Furthermore, the computational cost grows with the number of particles, limiting scalability to larger or more dynamic environments.
In contrast, our approach does not require complex modeling, as it leverages machine learning to learn positioning functions directly from measured data.

Measured CSI also enables neural positioning, which directly estimates UE positions with neural networks~\cite{Chapre2014CSIFingerprinting,He2016CSIFingerprinting,wang2017indoorLocalDeepLearning,wu2013csiIndoorLocal,Arnold2019CSIFingerprinting,Savic2015Figerprinting,Vieira2017Fingerprinting,Gönültaş2021probabilityFusion}.
Such methods enable highly-accurate off-device UE localization in complex indoor and outdoor scenarios. Nonetheless, training the positioning functions (the neural networks) typically requires ground-truth position labels obtained from an external reference positioning system (e.g., using a WorldViz precision position tracking system~\cite{worldviz} or a tachymeter~\cite{dichasus2021}).
In contrast, our approach eliminates this dependency, making it (i) inexpensive, (ii) easy to deploy, and (iii) possible to train (and retrain) neural positioning functions for large and complex environments. 

Channel Charting (CC) has recently emerged as a CSI-based neural pseudo-positioning method that avoids the need for ground-truth position labels altogether~\cite{studer2018cc}. In its original form, CC does not provide UE position information in real-world coordinates, but recent extensions enable UE positioning (e.g., by leveraging a digital twin~\cite{cazzella2025digitaltwin,mateosramos2025DTaidedChannelCharting} or AP position information~\cite{taner2025channel}). The positioning accuracy of such CC-based positioning methods, however, is significantly worse than supervised neural positioning methods~\cite{wiesmayr2025nvidia5gtestbed}.  

To mitigate the dependency on external reference positioning systems, several neural positioning methods have been developed that utilize internal sensor measurements (e.g., acquired on robot platforms that move through the area of interest), including data from inertial measurement units (IMUs)~\cite{Ermolov2023imupos,choi2022sensorAidedWiFiBeaconPos}. 
Reference~\cite{choi2022sensorAidedWiFiBeaconPos} proposes merging CSI- and RSSI-based UE position estimates with a sensor-reconstructed trajectory to enable on-device positioning. 
However, this method requires internal sensor data during position inference, is complex, and reports mean positioning errors exceeding $1$\,m.
In contrast, our approach only requires CSI data during position inference, is simple, leverages machine learning for positioning, and achieves centimeter-level accuracy.

Reference~\cite{Ermolov2023imupos} reconstructs absolute pseudo-labels through IMU-based trajectory integration, employing sophisticated IMU correction, drift compensation, and iterative refinement. 
While this approach trains accurate positioning functions, it suffers from three drawbacks. First, the pseudo-labels computation pipeline is complicated and complex. 
Second, positioning is performed on-device, preventing network providers from obtaining UE position information. 
Third, their real-world dataset captures UE movement only along a single L-shape trajectory.
In contrast, our method is simple, computationally efficient, and enables accurate off-device UE positioning.
Furthermore, we validated our approach across diverse real-world environments and with multiple wireless standards.

Similar ideas as in~\cite{Ermolov2023imupos,choi2022sensorAidedWiFiBeaconPos} have been developed for CC-based approaches~\cite{ahadi2025displacementCC,poeggel2025passivecclocating,euchner2025passivechannelchartinglocating}.
Reference~\cite{ahadi2025displacementCC} proposes a CC-based approach that utilizes TDoA information from 5G signals. An additional loss term based on displacement measurements between consecutive time stamps is used to promote geometric consistency with TDoA-based distances to multiple APs. 
Although this method avoids labeled ground-truth positions, it relies on potentially inaccurate TDoA measurements, requires accurate time synchronization, and is sensitive to multipath and NLoS conditions.
In contrast, our approach is insensitive to NLoS conditions and yields a simpler and more practical training pipeline.
References~\cite{poeggel2025passivecclocating,euchner2025passivechannelchartinglocating} analyze wireless signals that propagate through the environment and scatter off moving objects (e.g., a person holding a UE) in the environment. Such passive approaches struggle to perform positioning in the presence of multiple moving objects and achieve only poor accuracy. In contrast, our approach performs UE positioning from measured CSI,  eradicating the issue of multiple simultaneously present UEs and enabling orders-of-magnitude better positioning accuracy.

\subsection{Paper Outline}
The rest of this paper is organized as follows.
\fref{sec:ourmethod} proposes our neural positioning pipeline alongside the triangle displacement loss. 
\fref{sec:system} describes the neural positioning pipeline and discusses implementation aspects.
\fref{sec:testbeds_and_scenarios} presents the used 5G NR and IEEE 802.11 \WiFi testbeds and describes our measurement scenarios.  
\fref{sec:baselines} introduces baseline methods and \fref{sec:results} presents positioning results. 
\fref{sec:conclusions} concludes.

\subsection{Notation}
Non-boldface letters denote scalars (e.g., $a$ or $A$); boldface lowercase letters denote vectors (e.g., $\bma$) and uppercase letters denote matrices and higher-dimensional arrays (e.g., $\bA$).
The $A$-by-$A$ identity matrix is denoted by $\matidentity_A$, the $A$-dimensional all-zeros vector is denoted by $\mathbf{0}_A$, and the Euclidean norm of a vector $\bma$ is denoted by $\|\bma\|$.
Sets are denoted by uppercase calligraphic letters (e.g., $\setA$).
We use~$\mathfrak{L}$ to denote a generic loss function. Specific loss functions are indexed when needed (e.g., $\mathfrak{L}_{\textnormal{d}}$ stands for the displacement loss).
The $A$-dimensional multivariate Gaussian distribution with mean vector $\bma\in\reals^A$ and covariance matrix $\bA\in\reals^{A\times A}$ is denoted by $\mathcal{N}_A(\bma,\bA)$.

\section{Neural Positioning Without External Reference}
\label{sec:ourmethod}

We now describe our method for learning a neural positioning function that only requires internal (e.g., from the moving robot passing through the area of interest during the measurement campaign) displacement information, avoiding the need for an external reference positioning system. We start by outlining the underlying idea and then, propose a parametric extension as well as a novel triangle displacement loss, which we will use to learn our neural positioning functions.

\subsection{Operating Principle}
\label{sec:operatingprinciple}

We model the relation between two positions $\bmx_n\in\reals^D$ and $\bmx_{n+1}\in\reals^D$ in $D$-dimensional coordinates at consecutive discrete time indices $n$ and $n+1$, respectively, as follows:
\be
    \bmx_{n+1} = \bmx_n + \bmdelta_n.
    \label{eq:displacement}
\ee
Here, $\bmdelta_n = \bmx_{n+1} - \bmx_n$ is the \textit{true displacement} of the transmitting UE from time index $n$ to $n+1$. 
If one would know the initial position $\bmx_0$ and a complete series of~$N$ consecutive true displacements $\{\bmdelta_n\}_{n=0}^{N-1}$, then one could obtain all of the positions $\{\bmx_n\}_{n=0}^{N}$ by the recursion~\fref{eq:displacement}.
In practice, however, one only has access to noisy versions~$\tilde{\bmdelta}_n$ of the true displacements~$\bmdelta_n$, $n=0,\ldots,N-1$, 
which requires (i) a statistical model for the observation errors and (ii) a robust procedure for estimating the UE positions. 

In what follows, we model the observation process as 
\be
    \tilde{\bmdelta}_n = \bmdelta_n + \tilde{\bme}_n,
    \label{eq:displacementerrormodel}
\ee
where we assume the errors in the displacement measurement~$\tilde{\bmdelta}_n$ to be zero-mean  Gaussian $\tilde{\bme}_n \sim \mathcal{N}_D(\mathbf{0}_D, \tilde{E}_n\matidentity_D)$ and mutually independent over the time steps $n=0,1,2,\ldots$, with a time-step-dependent variance $\tilde{E}_n$. 

\begin{rem}
Time-step-dependent variances allow the error magnitude to depend on the displacement (e.g., larger displacements may suffer from larger errors); more complex (and more accurate) error models are possible but not pursued further. 
\end{rem}

\begin{rem}
The displacement measurement $\tilde{\bmdelta}_n$ in \fref{eq:displacementerrormodel} can either be a measurement of the displacement, e.g., extracted from an internal IMU, or simply the command provided to the robot that is performing the desired displacement. In what follows, we exclusively consider simple displacement commands.
\end{rem}

From \fref{eq:displacementerrormodel}, we can write the probability density function (PDF) representing the likelihood of the observed noisy measurement~$\tilde{\bmdelta}_n$ given the true displacement~$\bmdelta_n$ as
\be
f(\tilde{\bmdelta}_n \mid \bmdelta_n) 
= \frac{1}{(2\pi \tilde{E}_n)^\frac{D}{2}} \exp\!\left( -\frac{\| \tilde{\bmdelta}_n - \bmdelta_n \|^2}{2\tilde{E}_n} \right)\!. 
\ee
Thus, given $N$ displacement measurements $\{\tilde{\bmdelta}_{n}\}_{n=0}^{N-1}$ and assuming mutual independence among the relative displacements, the joint likelihood function is given by
\begin{align}
& f\!\left(\{\tilde{\bmdelta}_{n}\}_{n=0}^{N-1}\mid \{\bmdelta_{n}\}_{n=0}^{N-1}\right) \notag \\
& \quad = \prod_{n=0}^{N-1} f(\tilde{\bmdelta}_{n} \mid \bmdelta_{n}) 
 \,\propto\,
\exp\!\left( - \sum_{n=0}^{N-1}  
\frac{\| \tilde{\bmdelta}_{n} - \bmdelta_{n} \|^2}{2 \tilde{E}_n} \right)\!. \label{eq:displacementpdf}
\end{align}
By replacing $\bmdelta_n$ in \fref{eq:displacementpdf} by the position differences from \fref{eq:displacement}, taking the negative of the logarithm, and dropping constants, we obtain the following \emph{displacement loss}
\be
\mathfrak{L}_{\textnormal{d}}\!\left(\{\bmx_n\}_{n=0}^N\right)
= \sum_{n=0}^{N-1} 
\frac{\| \tilde{\bmdelta}_{n}- (\bmx_{n+1} - \bmx_{n}) \|^2}{2\tilde{E}_n},
\label{eq:max_likelihood}
\ee
which, given the set of  displacement measurements $\{\tilde{\bmdelta}_{n}\}_{n=0}^{N-1}$, could be minimized to attempt to estimate the UE transmit positions $\{\bmx_n\}_{n=0}^N$. This approach, however, is doomed to fail as we have $N+1$ UE positions and only $N$ displacement measurements, which---when minimized---would lead to position estimates up to an unknown global displacement. 

In order to address this nonuniqueness issue, one needs knowledge of at least one UE position---in practice, this is not an issue as one usually knows the initial position (e.g., the location of the charging dock).
As a result, we assume that one has access to a small set of $A$ estimated anchor positions $\{\tilde{\bmx}_a\}_{a\in\setA}$ with $\setA = \{1,2,\ldots,A\}$, which correspond to true anchor positions $\{\bmx_a\}_{a\in\setA}$ via the following error model
\be
\tilde{\bmx}_a = \bmx_a + \tilde{\bme}_a,
\ee
with the anchor position error $\tilde{\bme}_a \sim \normal_D( \mathbf{0}, 
\tilde{E}_a \matidentity_D)$. 
This error model gives rise to the following \emph{anchor loss}
\ba
\mathfrak{L}_{\textnormal{a}}\!\left(\set{\bmx_{a}}_{a \in \setA}\right)
& = \sum_{a \in \setA}
\frac{\| \tilde{\bmx}_{a} - \bmx_{a} \|^2}{2\tilde{E}_a}.
\label{eq:negativ_log_likelihood_anchor}
\ea

Consequently, by minimizing the sum of the displacement loss and anchor loss $\mathfrak{L}=\mathfrak{L}_{\textnormal{d}}+\mathfrak{L}_{\textnormal{a}}$, one could directly estimate the set of $N+1$ UE positions $\{\tilde{\bmx}_n\}_{n=0}^N$. 

\begin{rem}
We reiterate that only one anchor point (e.g., the position of the charging dock) is sufficient for obtaining a unique set of UE position estimates. 
\end{rem}

\begin{rem}
In absence of noise in both the displacement and the anchor measurements, the above estimation procedure would perfectly recover the true UE positions $\{\bmx_n\}_{n=0}^N$. 
\end{rem}

\subsection{Parametric Extension using Machine Learning}
The above procedure is nonparametric, i.e., given a set of displacement measurements and anchor positions, one can only produce a set of estimated UE positions related to the observed displacements---obtaining position estimates associated with new (previously unseen) UE positions from measured CSI, which is what is necessary for UE positioning, is not directly possible. We now introduce a parametric version of the above idea that relies on artificial neural networks. 

In what follows, we assume that we have a positioning function $\bmg_{\bmtheta}:\reals^F \to \reals^D$, with learnable parameters $\bmtheta$ (e.g., weights and biases of a neural network), that maps $F$-dimensional CSI features $\bmf_n\in\reals^F$ to position estimates in $D$ dimensions (typically $D=2$ or $D=3$) as follows:
\begin{align}
\bmg_{\bmtheta}(\bmf_n) = \est{\bmx}_n = \bmx_n + \bme_n.
\label{eq:positioningerror}
\end{align}
Here, $\bmx_n$ denotes the true position,  $\est{\bmx}_n$ the neural positioning output, and $\bme_n \sim \normal_D( \mathbf{0},E_n\matidentity_D)$ models the errors of the positioning function. For simplicity, we assume that these errors are independent and identically (i.i.d.) distributed with respect to the time index~$n$. Furthermore, we assume that $\tilde{\bme}_n$ is independent of~$\bme_n$.

\begin{rem}
Once again, more complex positioning function error models are possible but not pursued further. 
\end{rem}

We can now combine \fref{eq:positioningerror} with the displacement recursion in~\fref{eq:displacement} and follow the derivations in \fref{sec:operatingprinciple} to arrive at the following \emph{parametric displacement loss}
\be
\mathfrak{L}_{\textnormal{d}}\!\left(\bmtheta\right)
= \sum_{n=0}^{N-1} 
\frac{\|  \tilde{\bmdelta}_{n} - ( \bmg_{\bmtheta}(\bmf_{n+1}) - \bmg_{\bmtheta}(\bmf_{n}) ) \|^2}{2(\tilde{E}_n+2{E_n})}
\ee
and \emph{parametric anchor loss}
\ba
\mathfrak{L}_{\textnormal{a}}\!\left(\bmtheta\right)
& = \sum_{a \in \setA}
\frac{\| \tilde{\bmx}_{a} - \bmg_{\bmtheta}(\bmf_{a})  \|^2}{2(\tilde{E}_a+E_n)}.
\label{eq:anchorloss}
\ea
Both of these loss functions can be summed to a total loss function $\mathfrak{L}_{\textnormal{d}}(\bmtheta) + \mathfrak{L}_{\textnormal{a}}(\bmtheta)$,
which no longer depends on the estimated positions directly (in contrast to
\fref{eq:max_likelihood} and \fref{eq:negativ_log_likelihood_anchor}) and can be minimized, given a set of displacement measurements $\{\tilde{\bmdelta}_{n}\}_{n=0}^{N-1}$, estimated anchor positions $\{\tilde{\bmx}_a\}_{a\in\setA}$, as well as error variances $\{\tilde{E}_n,\tilde{E}_a,E_n\}$, to learn the function parameters~$\bmtheta$. 

\begin{rem}
Apart from a small set of anchor positions (remember: one anchor is sufficient), Eq.~\fref{eq:totalloss} enables learning of positioning functions by only requiring relative displacement information---this can be obtained without an external reference positioning system (e.g., with a robot that moves in an area by accepting a set of known displacement commands).
\end{rem}

\subsection{Improved Learning using a Triangle Displacement Loss}
\label{sec:triangle_loss}

The high rate of CSI acquisition in practical wireless systems results in a large number of small incremental displacements from one time instant to the next. Furthermore, most consecutive displacements point in the same direction, which (i) may enhance systematic displacement measurement errors and (ii) contradicts the assumption in \fref{sec:operatingprinciple} that the errors $\tilde{\bme}_n$ are mutually independent with respect to the time index $n$. 
If, however, a sufficiently large number of displacements is accumulated, the overall error becomes approximately Gaussian again---a consequence of the central limit theorem. 
As we will show in \fref{sec:results}, neural positioning function learning works significantly better with larger, accumulated  
displacements. 
Based on this insight, we propose a novel loss function called \emph{triangle displacement loss}, which improves learning.

\begin{figure}[tp]
    \centering
    \includegraphics[width=0.95\linewidth]{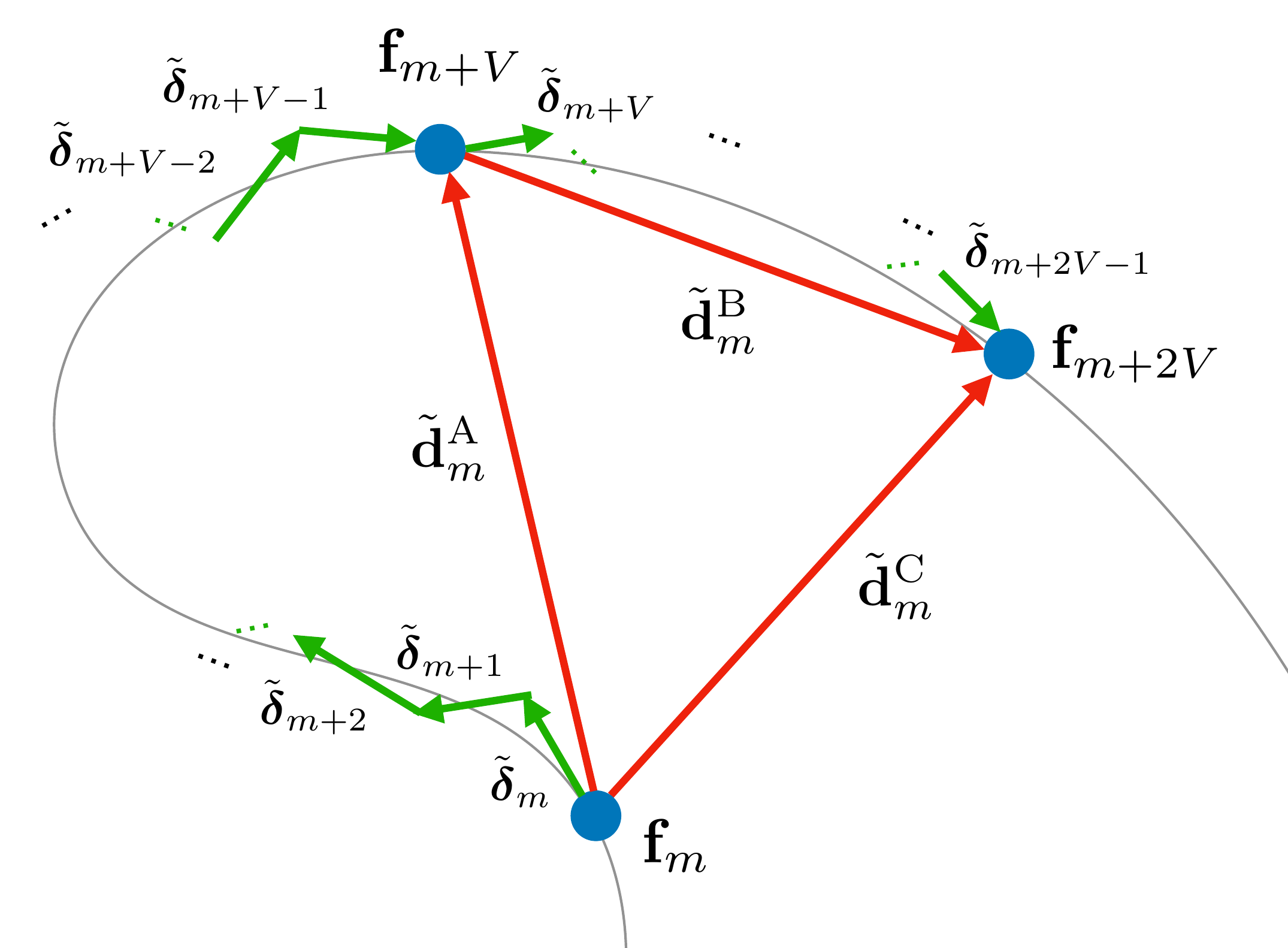}
    \vspace{-0.2cm}
    \caption{Illustration of the triangle displacement loss formed from a geometric triangle. Each vertex is associated with a CSI feature; arrows indicate the displacement vectors between endpoints used for training. The gray curve represents the true movement in space; the smaller green arrows indicate noisy displacement measurements.}
    \label{fig:triplet_triangle}
\end{figure}

The idea of the triangle displacement loss is to form triplets of endpoints, named \emph{triangle sets}, forming larger triangles in space. These triangles are constructed by accumulating multiple small relative displacements; see \fref{fig:triplet_triangle} for an illustration. 
For every triangle, with sides $\mathrm{A}$, $\mathrm{B}$, and $\mathrm{C}$, one picks a start time index $m$, which is associated with a CSI feature $\bmf_{m}$. 
To form the side $\mathrm{A}$ of the triangle (cf. \fref{fig:triplet_triangle}), one then computes a first large relative displacement as  
\begin{align}
\tilde{\bmd}^{\mathrm{A}}_{m} = \sum_{n=m}^{m+V-1} \tilde{\bmdelta}_{n},
\label{eq:delta_A}
\end{align}
where $V$ is a given (and fixed) \emph{leap increment}\footnote{We study the impact on positioning accuracy depending on the leap increment $V$ in \fref{sec:results}.}. The vertex index $m+V$ is associated with CSI feature $\bmf_{m+V}$.
One then computes a second large relative displacement as
\begin{align}
\tilde{\bmd}^{\mathrm{B}}_{m} = \sum_{n=m+V}^{m+2V-1} \tilde{\bmdelta}_{n},
\label{eq:delta_B}
\end{align}
which forms the side $\mathrm{B}$ with endpoint associated to CSI feature~$\bmf_{m+2V}$.
Finally, we form the displacement of side~$\mathrm{C}$ (notice the direction of the arrows in \fref{fig:triplet_triangle}) simply by completing the triangle as follows:
\begin{align}
\tilde{\bmd}^{\mathrm{C}}_{m} = \tilde{\bmd}^{\mathrm{A}}_{m} + \tilde{\bmd}^{\mathrm{B}}_{m}.
\label{eq:delta_C}
\end{align}

From this procedure, one can form the following loss for each triangle $\mathrm{A}\mathrm{B}\mathrm{C}$ starting at sample index $m$:
\begin{align}
\mathfrak{L}_{\textnormal{t},m}(\bmtheta) 
= \,\, & 
\frac{\| \tilde{\bmd}^{\mathrm{A}}_{m} - (\bmg_\bmtheta(\bmf_{m+V}) - \bmg_\bmtheta(\bmf_{m}))  \|^2}{2E_{m}^{\mathrm{A}}} \\
& +  \frac{\| \tilde{\bmd}^{\mathrm{B}}_{m} -( \bmg_\bmtheta(\bmf_{m+2V}) - \bmg_\bmtheta(\bmf_{m+V}) ) \|^2}{2E_{m}^{\mathrm{B}}} \\
& + \frac{\| \tilde{\bmd}^{\mathrm{C}}_{m} - (\bmg_\bmtheta(\bmf_{m+2V}) - \bmg_\bmtheta(\bmf_{m})) \|^2}{2E_{m}^{\mathrm{C}}}.
\end{align}
Note that here each side of the triangle is associated with a specific displacement error variance ($E^\mathrm{A}_m$, $E^\mathrm{B}_m$, and $E^\mathrm{C}_m$).
These depend on the error variances $\tilde{E}_n$ associated with the small displacements that were used to make up the triangle's sides, as well as on the positioning function’s variance $E_n$.

The final \emph{parametric triangle displacement loss} is then  given by summing all of the individual triangle losses:
\begin{align}
\mathfrak{L}_{\textnormal{t}}(\bmtheta) = \sum_{m=0}^{N-2V}\mathfrak{L}_{\textnormal{t},m}(\bmtheta).
\label{eq:triangle_loss}
\end{align}
Together with the parametric anchor loss in \fref{eq:anchorloss}, we propose to minimize the following total loss function
\begin{align}
\mathfrak{L}(\bmtheta)
= \mathfrak{L}_{\textnormal{t}}(\bmtheta) + \mathfrak{L}_{\textnormal{a}}(\bmtheta),
\label{eq:totalloss}
\end{align}
which enables one to learn the neural positioning function~$\bmg_{\bmtheta}$ that maps CSI features to estimated UE position.

\section{System Implementation}
\label{sec:system}

We now provide an end-to-end description of the neural positioning pipeline based on the method proposed in \fref{sec:ourmethod}. We first give an overview, followed by a description of the process of transforming CSI and displacement measurements into CSI features and displacement vectors. We conclude this section by detailing our machine-learning method that yields the neural positioning function.

\begin{figure*}[t]
  \centering
  \subfloat[]{
    \includegraphics[width=0.40\textwidth]{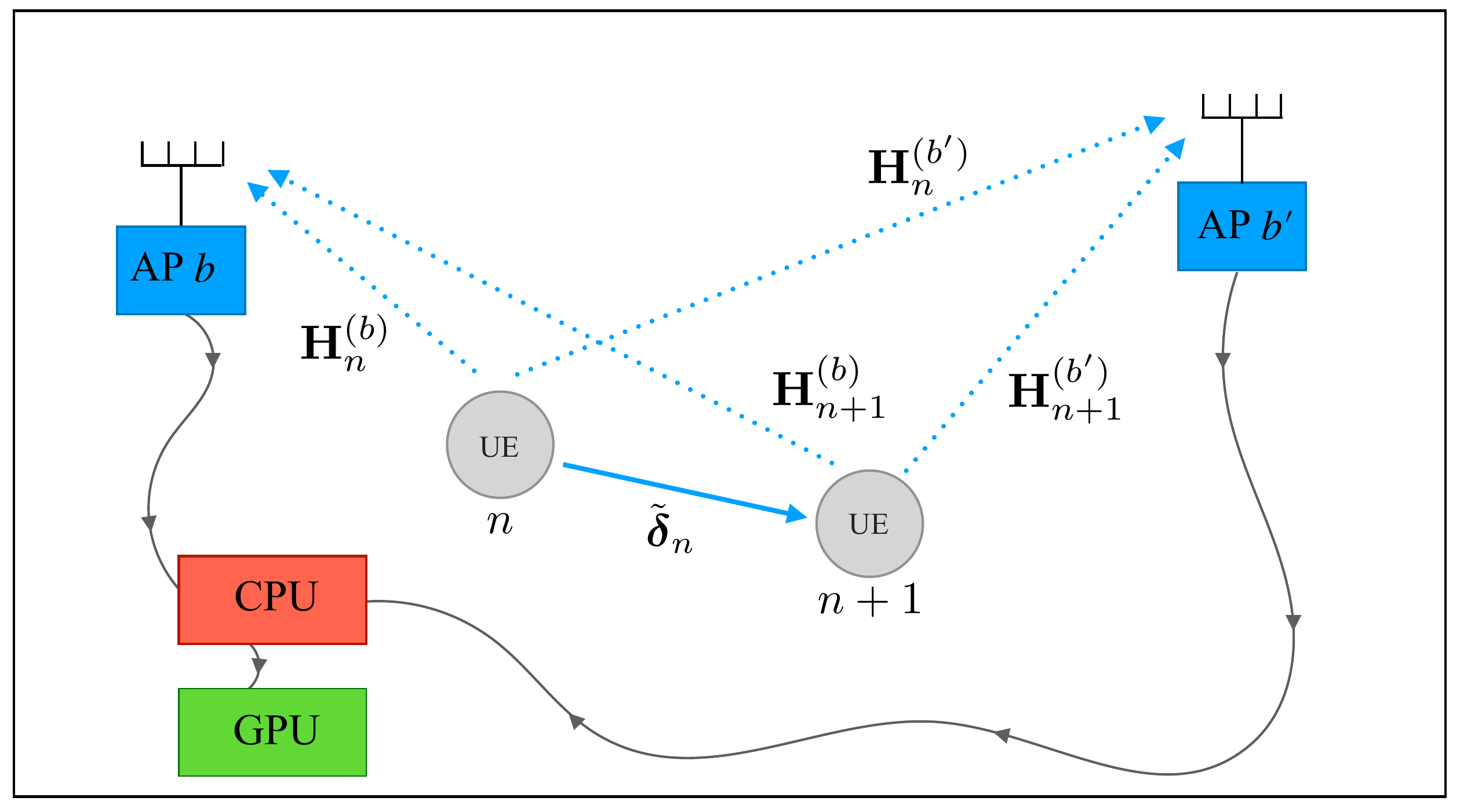}%
    \label{fig:Measurement_Sketch}%
  }\hfill
  \subfloat[]{
    \includegraphics[width=0.58\textwidth]{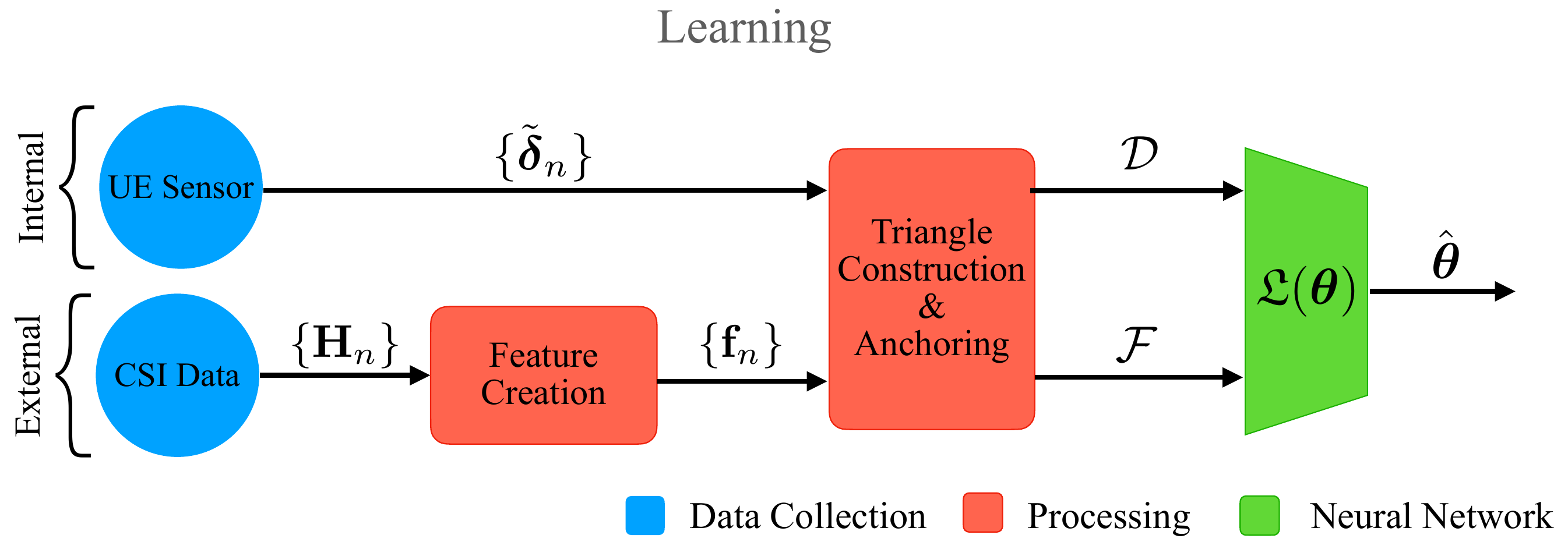}%
    \label{fig:ML_Pipline_Sketch}%
  }
  \caption{Overview of the UE positioning pipeline: (a) Measurement setup: A UE moves from time stamp $n$ to $n+1$ and records the displacement measurement~$\tilde{\bmdelta}_n$ (in our case, corresponding to the executed displacement command). Distributed receive APs $b,b' \in \{1,\dots,B\}$ collect the CSI at each time stamp, which is preprocessed on a CPU and then used for model training on a GPU. (b) Neural positioning function learning: CSI is transformed into CSI features and relative displacement data is recorded. The CSI features and displacement measurements are grouped into sets of triangle vertices to form the collection of CSI features~$\setF$ and large relative displacements~$\setD$. Finally, the neural positioning function parameters are trained using the triangle displacement loss.}
  \label{fig:feature_processing_overview}
\end{figure*}

\subsection{System Overview}
The proposed neural positioning system is illustrated in Fig.~\ref{fig:feature_processing_overview}. 
First, CSI data of the wireless channels between the transmitting UE and the receiving APs is collected at the APs---simultaneously, displacement measurements (i.e., the robot's displacement commands) are recorded internally at the UE.
Although not required for the proposed position estimation method, ground-truth position data of the UE is collected solely for the purpose of allowing a precise evaluation of the estimated positions against the true trajectory.

\begin{rem}
Our neural positioning method does \emph{not} require external ground-truth position measurements. We record ground-truth positions \emph{only} to assess positioning accuracy.
\end{rem}

After data collection, the following steps are executed for neural position function learning.
First, the measured CSI data is transformed into CSI feature vectors to minimize small-scale fading effects \cite{studer2018cc,gönültas2021featuresForCSIpositioning,mateosramos2025DTaidedChannelCharting}. 
Second, the resulting CSI feature vectors together with the estimated UE displacement vectors are grouped into corresponding triangle sets, each representing a triangle in space (cf.~\fref{sec:triangle_loss}). In addition, a small set~$\setA$ of anchor features is identified---we only use the location of the charging dock of the robot platform as anchor.
Third, the triangle sets as well as the anchor set are used for training a machine-learning model, resulting in the estimated model parameters $\est{\bmtheta}$.
The positioning function $\bmg_{\bmtheta}:\reals^F \to \reals^D$ to be learned is realized by a simple  multilayer perceptron (MLP); see \fref{sec:ML_pipeline} for the details. 

During the UE positioning phase, a test set of CSI features $\{\bmf_{n'}\}_{n'=0}^{N'}$ that were excluded in the training phase are fed into the trained model~$\bmg_{\est{\bmtheta}}$ to obtain position estimates $\{\est{\bmx}_{n'}\}_{n'=0}^{N'}$. These estimated positions are then compared to the recorded ground-truth positions to assess positioning accuracy.

\subsection{CSI Data and CSI Features}
\label{sec:CSI_Preprocessing}

The CSI data used as external inputs to our system are acquired in distributed single-input multiple-output (SIMO) systems, where at time stamp $n$ one UE transmits pilots to several multi-antenna APs using orthogonal frequency-division multiplexing (OFDM).
Thus, each CSI datapoint $\bH_n\in \mathbb{C}^{B \times A \times W}$ at time stamp $n$ corresponds to a three-dimensional array formed by $B$ APs each with $A$ antennas and $W$ active subcarriers.
In what follows, we acquire CSI data for $N+1$ time stamps $n=0,1,\ldots,N$ and classify the complete CSI dataset $\{\bH_n\}_{n=0}^N$ as \textit{external} data.

\begin{rem}
In this work, our CSI datasets are measured with two different systems: an IEEE 802.11 \WiFi testbed and a 5G NR testbed. Both of these testbeds are detailed in \fref{sec:testbeds}.
\end{rem}

Before using these CSI measurements for training, the CSI data $\{\bH_n\}_{n=0}^N$ is preprocessed into CSI features analogously to the work in \cite{zumegen2024wifiCSI}.
First, for every time stamp $n=0,1,\ldots,N$, we compute the magnitude of each complex-valued subcarrier, discarding phase information. 
Second, we normalize the magnitudes to unit Euclidean norm over all $B$ APs and $A$ receive antennas.
Third, we vectorize the resulting data into a CSI vector $\bmf_n \in \mathbb{R}^{F}$ of dimension $F$. 
The resulting CSI feature dataset $\set{\bmf_n}_{n=0}^N$ is then passed to the meachine-learning pipeline described in \fref{sec:ML_pipeline}.

\subsection{Displacement Data Preprocessing}
\label{sec:displacementdataproc}

We classify the measured displacement data in our system as \textit{internal} data, as it is extracted locally at the UE from simple displacement commands provided to the robot platform.
The measured two-dimensional displacements $\{\tilde{\bmdelta}_{n}\}_{n=0}^{N-1}$ are processed into large displacement vectors according to \fref{eq:delta_A}, \fref{eq:delta_B}, and \fref{eq:delta_C} to form the triangle sides $\mathrm{A}$, $\mathrm{B}$, and $\mathrm{C}$.
The resulting triangle sides are grouped as triangle sets to form the triangle dataset
\be \label{eq:triangledatasets}
    \setD = \set{(\tilde{\bmd}_{m}^{\mathrm{A}}, \tilde{\bmd}_{m}^{\mathrm{B}}, \tilde{\bmd}_{m}^{\mathrm{C}}) : m = 0, \ldots, N - 2V }\!,
\ee
which is then fed into the learning pipeline described next.

\begin{rem}
The displacement commands given to our mobile robot platform (described in \fref{sec:robotplatform}) exhibit a systematic bias with respect to the true traveled distance. We once measure this bias and then subtract it from the displacement commands.
\end{rem}

\subsection{Learning and Positioning Pipeline}
\label{sec:ML_pipeline}

For neural positioning, we use a five-layer fully-connected  MLP that is trained using the CSI features from \fref{sec:CSI_Preprocessing} and the large displacement vectors from \fref{sec:displacementdataproc}. 
The MLP consists of an input layer, three hidden layers, and an output layer, with the architecture details specified in \fref{tbl:NN_parameters}. All hidden layers use ReLU activations; the output activation, which produces UE position estimates, is linear.

\begin{table}[t]
\centering
\caption{Neural positioning function architecture.}
\label{tbl:NN_parameters}
\renewcommand{\arraystretch}{1.4} 
\begin{tabular}{@{}ll@{}}
\toprule
\textbf{Parameter} & \textbf{Description} \\ \midrule
Architecture type & Fully-connected multilayer perceptron (MLP) \\
Number of layers   & Input, $3$ hidden layers, output \\
Layer dimensions   & $F$, 512, 256, 64, 2 \\
Activation functions & ReLU (hidden layers), linear (output) \\
\bottomrule
\end{tabular}
\end{table}

\subsubsection{Positioning Function Learning}
First, the CSI features $\set{\bmf_n}_{n=0}^N$ are grouped into triangle sets in the collection
\be
    \setF = \set{(\bmf_{m}, \bmf_{m+V}, \bmf_{m+2V}) : m = 0, \ldots, N - 2V},
\label{eq:feature_collection}
\ee
thereby resembling the same triangles as the collection of large displacement vectors ${\mathcal{D}}$ in \fref{eq:triangledatasets}.
Then, the inputs to the MLP training algorithm, i.e., the training set, are (i) the elements of the displacement vector collection $\mathcal{D}$, (ii) the elements of the corresponding CSI feature collection $\setF$, and (iii) the set of anchored features $\{\bmf_n\}_{n \in \setA}$ and the associated estimated anchor positions $\{\tilde{\bmx}_a\}_{a\in\setA}$.

For learning the  parameters~$\hat{\boldsymbol{\theta}}$ of the positioning function $\bmg_{\hat{\boldsymbol{\theta}}}$, we utilize the total loss function~\fref{eq:totalloss} and set all of the involved error variances $E^\mathrm{A}_m$, $E^\mathrm{B}_m$, and $E^\mathrm{C}_m$ for $m=0,\ldots,N-2V$, and $\tilde{E}_a$ for $a\in\setA$ to one. 

\begin{rem}
The use of accurate error-variance values that depend, e.g., on instantaneous displacements and/or the utilized robot platform, might further improve UE positioning accuracy. This is, however, left for future work, mainly, as the achieved positioning accuracy is already high; see \fref{sec:results} for the results. 
\end{rem}

We use PyTorch~\cite{paszke2019pytorch} with the Adam optimizer~\cite{kingma2015adam} and an initial learning rate of $10^{-4}$ with step-size decay.
For all experiments (including  baselines), we carry out $15$ training epochs; each epoch corresponds to a complete pass over all datapoints in the respective training set. 
Across all experiments and scenarios, we split the measured data into training and test sets using a randomly sub-sampled 4:1 train-to-test ratio.
We use an \emph{NVIDIA GeForce RTX 4070} GPU for both training and inference. The proposed neural positioning pipeline---from feature creation to neural network learning---takes less than $10$ and $35$ minutes for the \WiFi and 5G NR datasets, respectively.

\subsubsection{Neural Positioning (Inference)}
For testing the learned positioning function $\bmg_{\hat{\boldsymbol{\theta}}}$, we use a separate test set of CSI features $\set{\bmf_{n'}}_{n'=0}^{N'}$ for which we compute $\est{\bmx}_{n'} = \bmg_{\est{\bmtheta}}(\bmf_{n'})$, $n'=0,\ldots,N'$, where $N'+1$ equals the number of test samples.
For testing, no displacement information is required.

\section{Testbeds and Scenarios}
\label{sec:testbeds_and_scenarios}

We now detail the two testbeds used for acquiring real-world CSI measurements and the three scenarios used to evaluate the proposed neural positioning pipeline. We also provide details on a \WiFi-testbed–specific CSI feature processing step that is necessary to attain accurate UE positioning.

\begin{figure*}[t]
  \centering

  \subfloat[]{
    \includegraphics[width=0.32\textwidth]{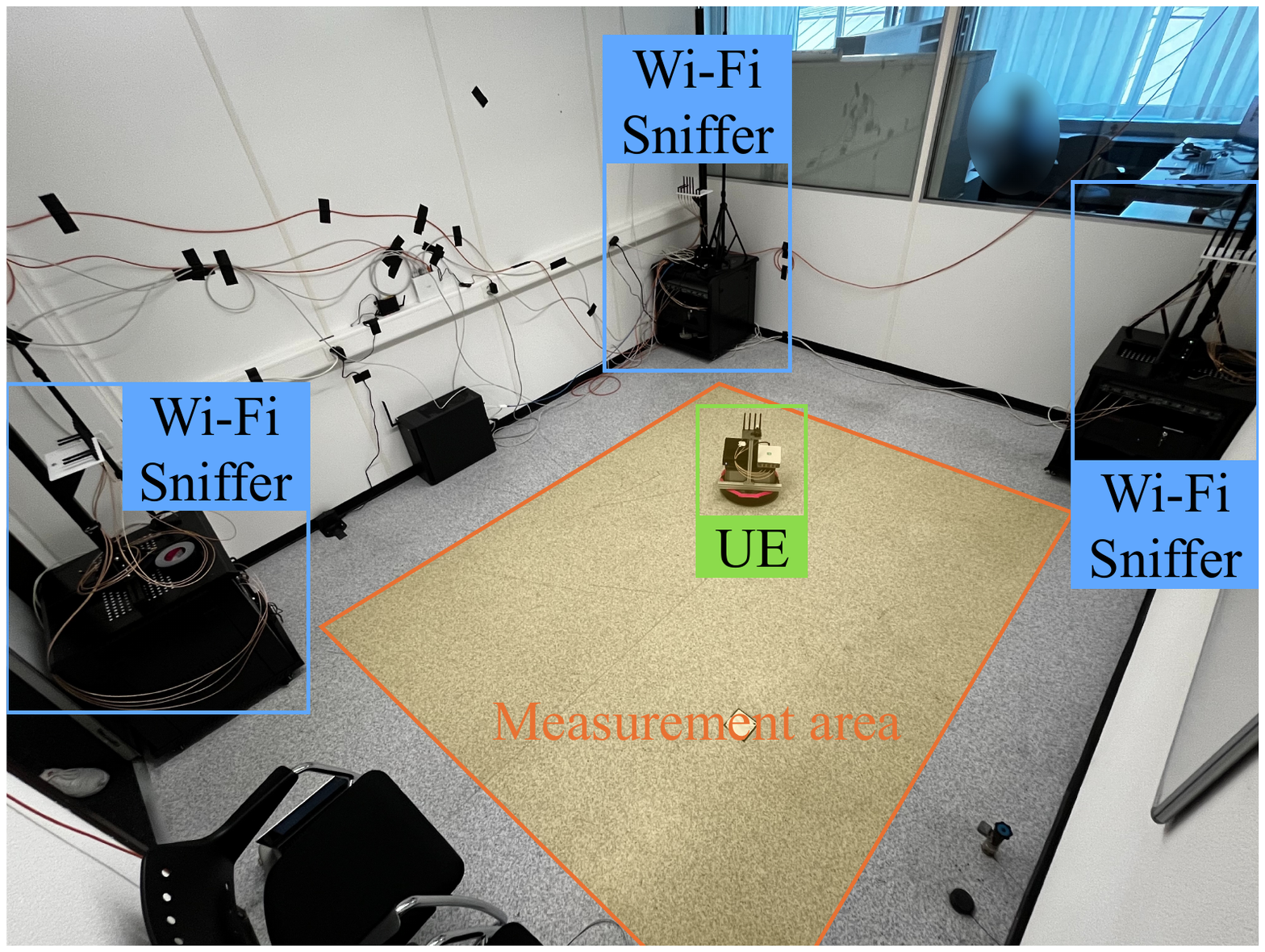}%
  }\hfill
  \subfloat[]{
    \includegraphics[width=0.32\textwidth]{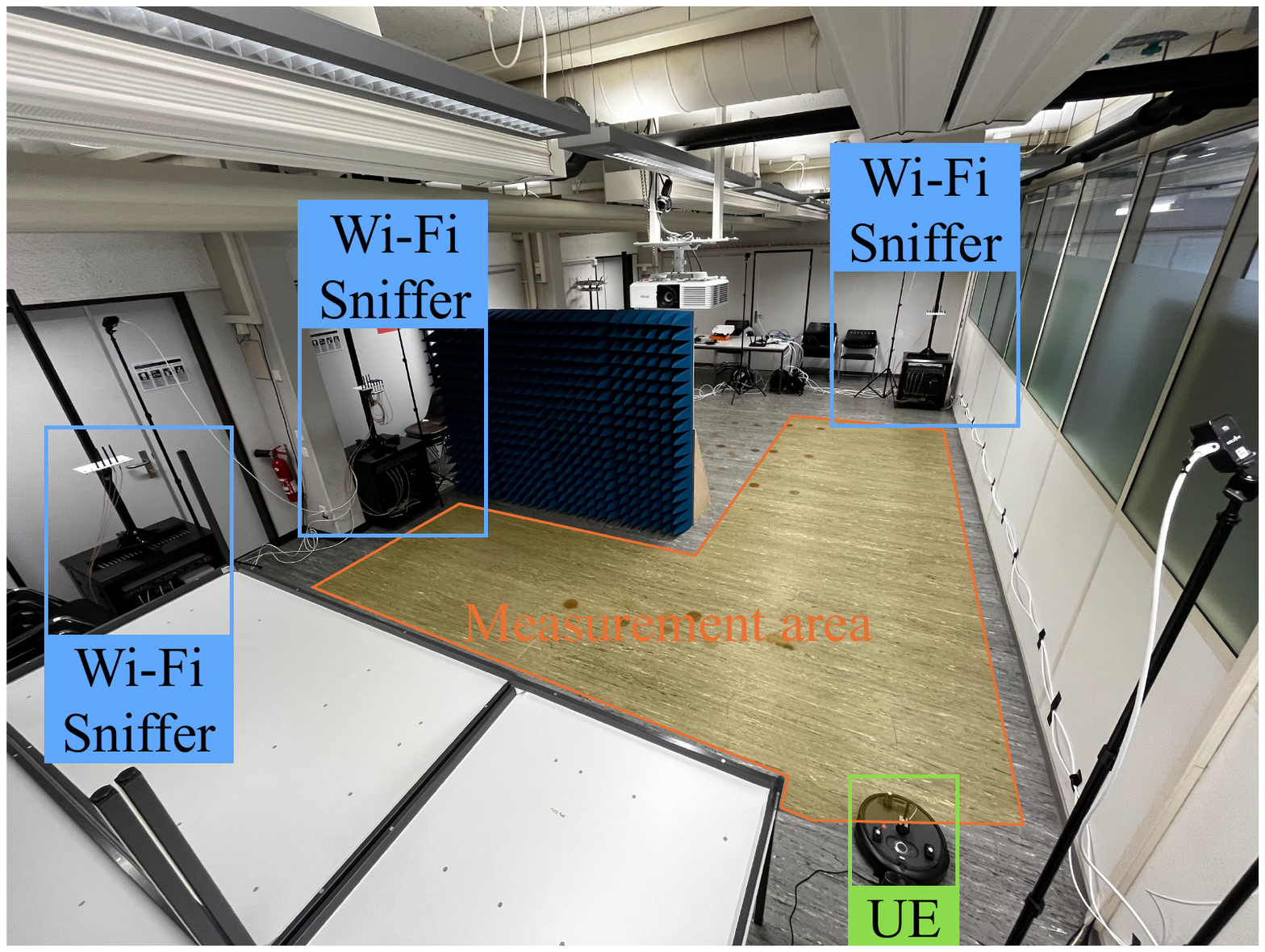}%
  }\hfill
  \subfloat[]{
    \includegraphics[width=0.32\textwidth]{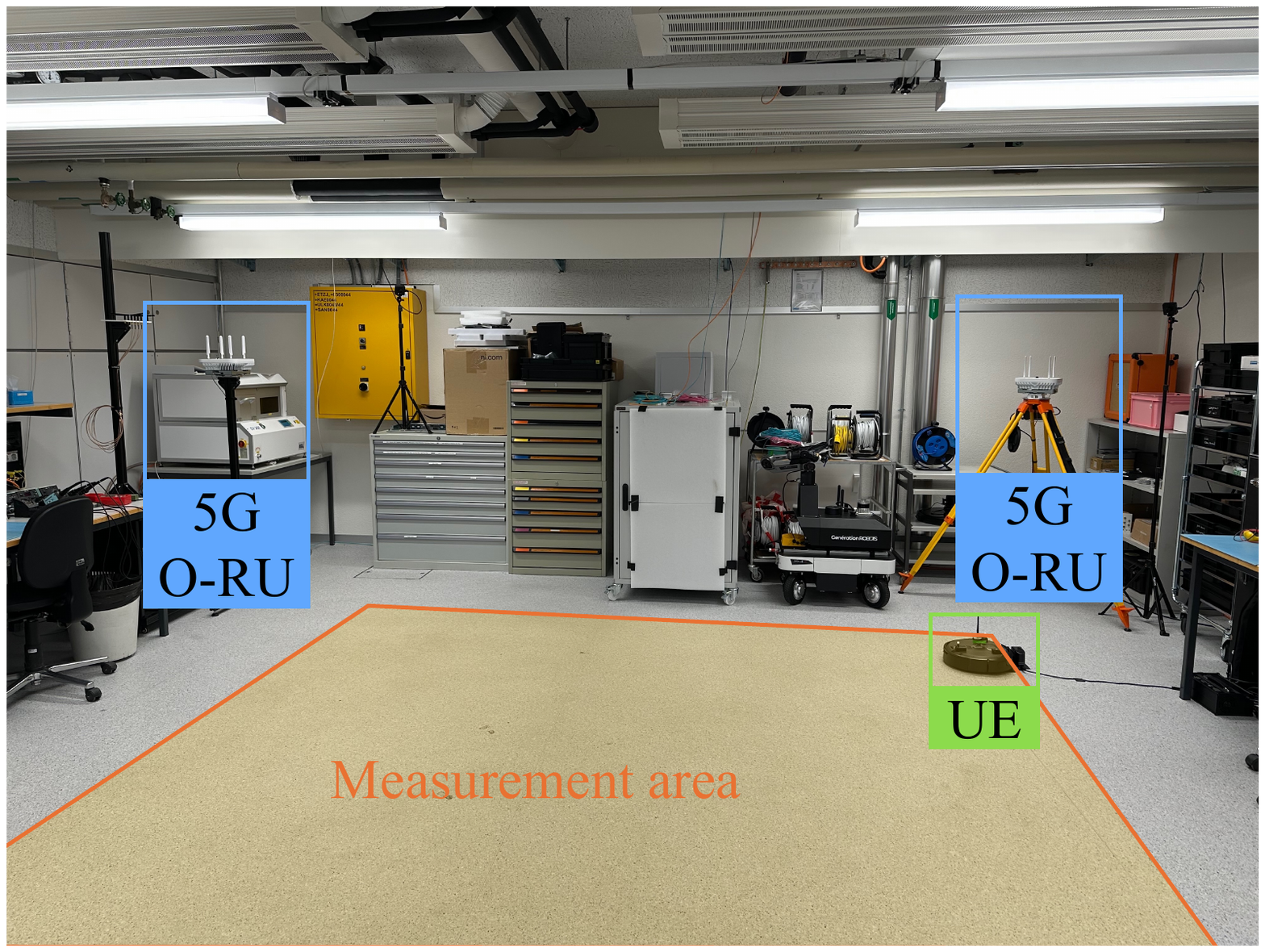}%
  }

  \vspace{0.5em} 

  \subfloat[]{
    \includegraphics[width=0.32\textwidth]{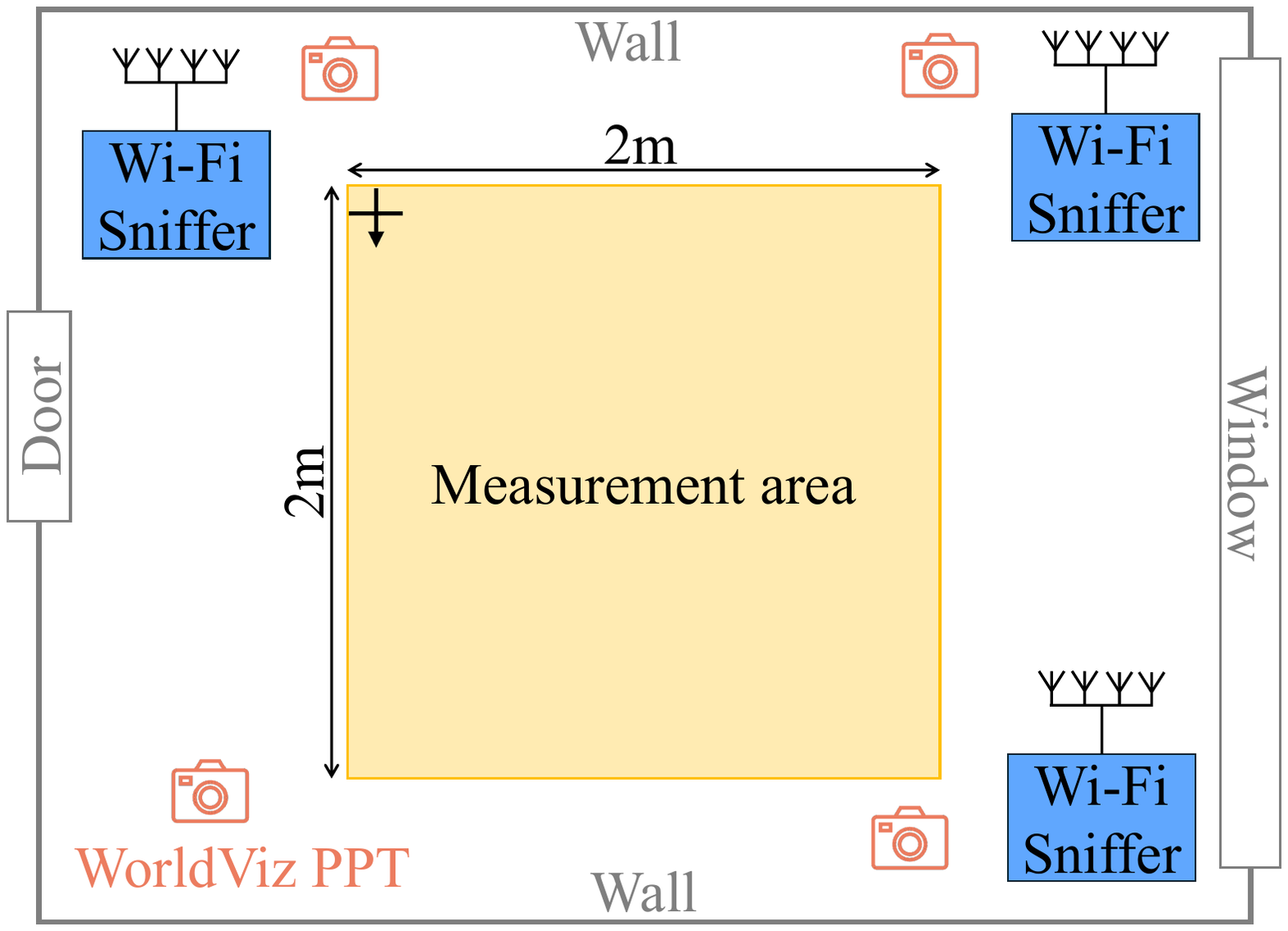}%
  }\hfill
  \subfloat[]{
    \includegraphics[width=0.32\textwidth]{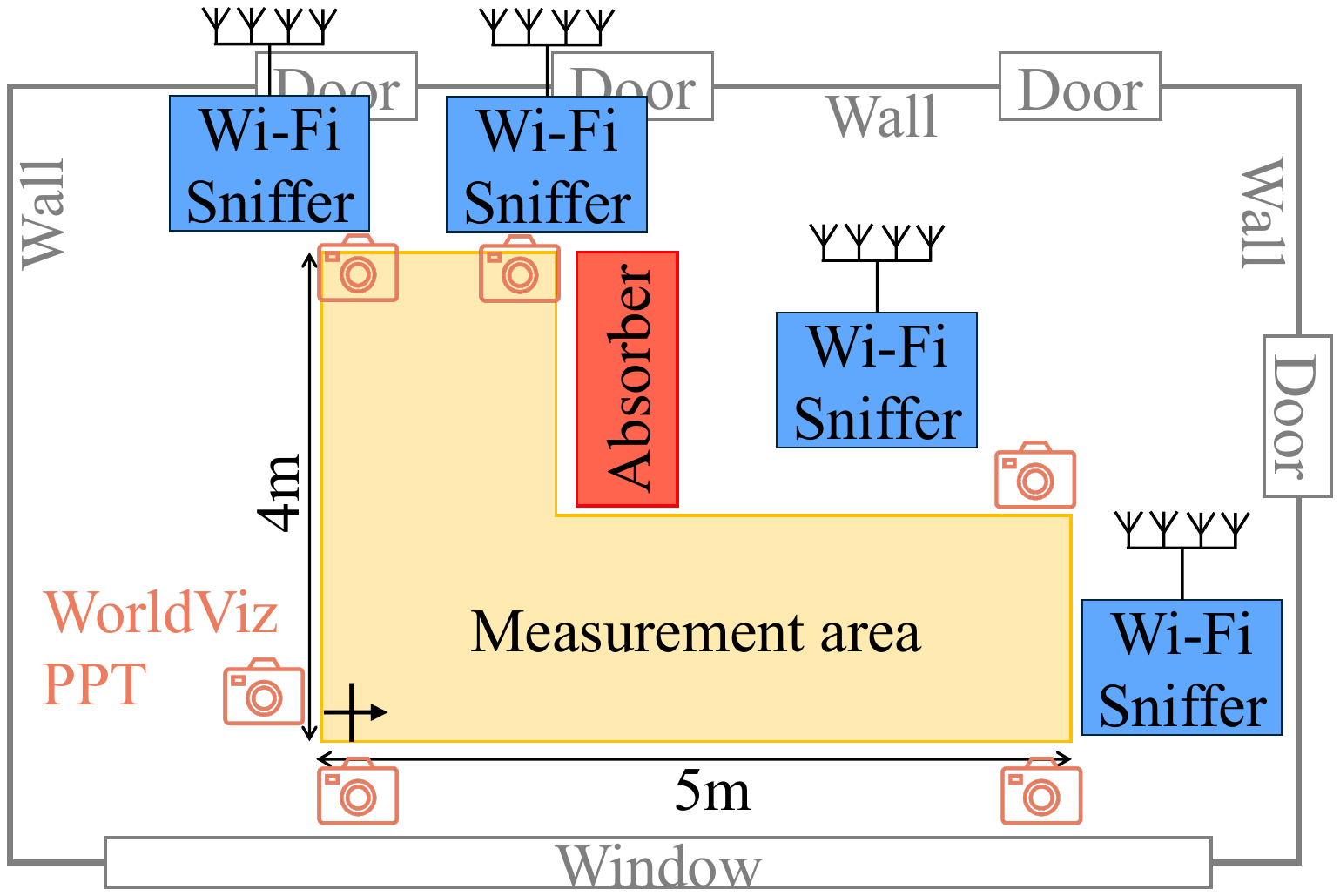}%
  }\hfill
  \subfloat[]{
    \includegraphics[width=0.32\textwidth]{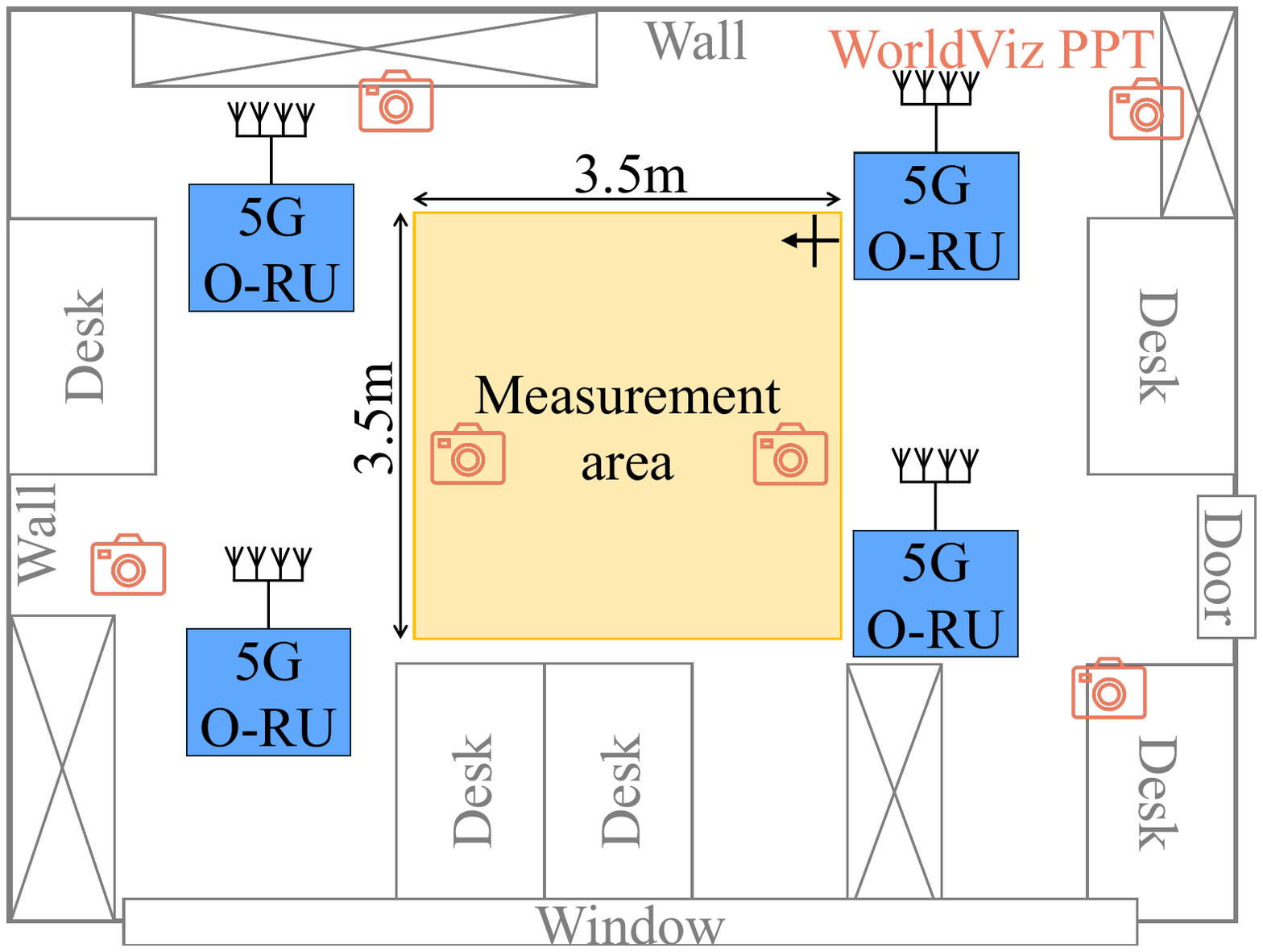}%
  }

  \caption{\WiFi and 5G testbeds in three scenarios: (a) and (d) show the small meeting room \WiFi scenario, (b) and (e) the large meeting room \WiFi scenario with partial NLoS, and (c) and (f) show the 5G office scenario with human activity. The top row shows photos of each scenario; the bottom row shows the respective floorplans. The UE platform's dock position is denoted by a cross with an arrow indicating the initial orientation.}
  \label{fig:testbeds}
\end{figure*}

\subsection{Testbeds and Robot Platform}
\label{sec:testbeds}

The CSI data used in this work is acquired from two different testbeds deployed at ETH Zurich: a \WiFi testbed and a 5G NR testbed. 
Both testbeds are set up in several indoor environments---what we call \emph{scenarios}---and are used for real-world CSI data collection. 
In each scenario, a transmitting UE is mounted onto a COTS robot platform that traverses through a predefined measurement area in randomized fashion following random displacement commands---those commands correspond to the internally-obtained displacement measurements.

\subsubsection{\WiFi Testbed} The \WiFi testbed uses multiple \WiFi sniffers as receivers to measure CSI~\cite{zumegen2024wifiCSI}. Each \WiFi sniffer utilizes four-antenna software-defined radios (SDRs) and a custom PHY-layer software stack that decodes \WiFi traffic. The \WiFi traffic is generated by a UE communicating in an active \WiFi network. Based on the extracted MAC address from a \WiFi frame, each sniffer determines whether the estimated CSI from that frame belongs to the channel between the UE and the sniffer's Rx array. The resulting CSI estimates are stored locally on the sniffer's host PC and processed later to create a combined CSI dataset for all sniffers. The key OFDM parameters of the \WiFi testbed are summarized in \fref{tbl:ofdm_parameter}; more details can be found in~\cite{zumegen2024wifiCSI}

\subsubsection{5G NR Testbed} The 5G NR testbed implements a software-defined full-stack 5G NR system and acquires CSI from the 5G NR PUSCH~\cite{wiesmayr2025nvidia5gtestbed}. The 5G NR testbed comprises four O-RUs with four antennas each. The software-defined NVIDIA Aerial physical layer estimates the uplink CSI to each of the four O-RUs. The 5G UE is only served by one O-RU, while the other three O-RUs are passive listeners. The software-defined MAC layer schedules the UE's PUSCH at least every $20$\,ms with always the full bandwidth of $100$\,MHz. The UE's transmit power is controlled by an outer feedback loop with a target SNR of $28$\,dB at the serving O-RU. The CSI estimates from all four O-RUs are stored in a database on the basestation's host server.
The key OFDM parameters of the 5G NR testbed are summarized in \fref{tbl:ofdm_parameter}; more details can be found in~\cite{wiesmayr2025nvidia5gtestbed}.

\begin{table}[t]
\centering
\caption{Key OFDM system parameters of the two testbeds.}
\renewcommand{\arraystretch}{1.4} 
\begin{tabular}{@{}lcc@{}}
\toprule
\textbf{Parameter} & \textbf{\WiFi Testbed} & \textbf{5G Testbed} \\
\midrule
Wireless standard & IEEE 802.11a & 5G NR \\
Carrier frequency & 5.18\,GHz & 3.45\,GHz \\
Bandwidth & 20\,MHz & 100\,MHz \\
Active subcarriers $W$ & 52 & 3'276 \\
Subcarrier spacing & 312.5\,kHz & 30\,kHz \\
\bottomrule
\end{tabular}
\label{tbl:ofdm_parameter}
\end{table}

\subsubsection{Robot Platform}
\label{sec:robotplatform}
Both testbeds receive data from a single-antenna UE mounted on top of a mobile robot platform. The platform used in our experiments is the inexpensive \emph{iRobot Create 3}~\cite{iRobot_Create3}, which is based on a popular COTS vacuum-cleaning robot.
The mobile platform with the mounted UE is controlled by a Python script and a ROS2 architecture \cite{macenski2022ros2}.
The Python script sends displacement commands to ROS2 nodes running on the mobile platform.
The displacement commands alternate between driving a predefined distance in the forward direction and rotating by a predefined angle left or right.
These displacement commands are drawn from independent uniform random distributions and correspond, after compensating for the systematic bias, to the displacement measurements $\{\tilde{\bmdelta}_n\}_{n=0}^{N-1}$.

\subsection{Measured Scenarios}
\label{sec:measured_scenarios}

To validate our proposed neural positioning pipeline, we conduct real-world experiments in three different scenarios: (i) a small meeting room, (ii) a large meeting room, and (iii) an office with human activity.
Each scenario was measured using one of the testbeds described in \fref{sec:testbeds} to capture CSI and displacement measurements.
The \WiFi testbed is used for scenarios (i) and (ii); the 5G NR testbed is used for scenario (iii).
See \fref{fig:testbeds} for the details on the three scenarios. 

Evaluating the accuracy of the proposed neural positioning pipeline requires a ground-truth positioning reference. To this end, all scenarios were equipped with the camera-based reference positioning system WorldViz PPT~\cite{worldviz}. This system provides ground-truth position information with sub-millimeter accuracy.
We remind the reader that the recorded reference positions are \emph{not} used for training our proposed method.

Each scenario contains a predefined measurement area to which the robot's movement is confined.
In all three scenarios, the robot starts from its charging dock in one corner of the measurement area (see \fref{fig:testbeds}). The dock positions are known and serve as the only anchor points used in our experiments.
The robot randomly traverses the measurement area and regularly returns to the charging dock.

The following paragraphs provide scenario-specific details. For each scenario, we generated one dataset. The details of these datasets are given in \fref{tbl:datasets}.
In the datasets and all following paragraphs, we refer to the \WiFi sniffers and the 5G O-RUs as \emph{APs} for simplicity.
For each sample $n = 0,1,\dots, N$ in these datasets, we recorded a ground-truth UE position using a WorldViz PPT reference positioning system.
One more time: We emphasize that the recorded reference positions are never used for training of our proposed method.

\subsubsection{\WiFi Small Meeting Room}
The measurement area spans a rectangular area of approximately $2$\,m$\times$$2$\,m. We placed three \WiFi sniffers in the corners of the room. The utilized single-antenna UE is custom-built based on an SDR.

\subsubsection{\WiFi Large Meeting Room}
The measurement area is L-shaped with outer side-lengths of $4$\,m$\times$$5$\,m. We placed four \WiFi sniffers outside the measurement area. To create partial NLoS conditions for at least one sniffer at a time with respect to any position inside the measurement area, we placed a wall of RF absorbers on the inner side of the L-shape. The utilized single-antenna UE is a COTS USB \WiFi adapter.
\subsubsection{5G Office with Human Activity}
The measurement area spans a rectangular area of approximately $3.5$\,m$\times$$3.5$\,m. We placed four 5G NR O-RUs outside the measurement area. At least one person (i.e., the measurement operator) was present in the office during dataset collection. The utilized single-antenna UE is a COTS 5G modem.

\begin{table}[t]
\caption{Summary of measured CSI datasets.}
\label{tbl:datasets}
\begin{minipage}{0.99\columnwidth}
\centering
\renewcommand{\arraystretch}{1.4} 
\begin{tabular}{@{}lccc@{}}
\toprule
& \textbf{\WiFi Small}  & \textbf{\WiFi Large}  & \textbf{5G} \\
& \textbf{Meeting Room} & \textbf{Meeting Room} & \textbf{Office} \\
\midrule
Measurement duration & 4\,h 0\,min & 3\,h 10~\,min & 2\,h 0\,min \\
Number of samples & 300'000 & 113'771 & 407'730 \\
APs\footnote{\WiFi sniffers and 5G O-RUs are both refered to as APs.} $B$ & 3 & 4 & 4 \\
Antennas/AP $A$ & 4 & 4 & 4 \\
Feature dimension $F$ & $B\!\cdot\!A\!\cdot\!W$ & $B\!\cdot\!A\!\cdot\!W$ & $B\!\cdot\!A\cdot\,$273\footnote{The CSI absolute values from the 5G NR system are down-sampled by a factor of $12$ in the subcarrier domain to reduce the large feature dimension.}\\
\bottomrule
\end{tabular}
\end{minipage}
\end{table}

\subsection{\WiFi CSI Feature Averaging}
\label{sec:wifi_feature_averaging}

The CSI features acquired through the \WiFi testbed suffer from several hardware impairments that require an additional postprocessing step to deliver accurate positioning performance. To this end, we apply a moving-average filter on the CSI feature vectors $\{\bmf_n\}_{n=0}^N$ with a window of length $L_n+1$. Specifically, we compute averaged feature vectors (for an even $L_n$) as
\begin{align}\label{eq:moving_average}
    \tilde{\bmf}_n = \frac{1}{L_n + 1} \sum_{\ell=n-L_n/2}^{n+L_n/2} \bmf_{\ell} \quad n=0,1,\ldots,N,
\end{align}
where we set $\bmf_{\ell} = \mathbf{0}_F$ for $\ell < 0$ and $\ell > N$.

In~\fref{sec:impact_of_L}, we evaluate several values for $L_n$. Generally, we either (i) set $L_n = L$ to a fixed value for all $n$, or (ii) determine $L_n$ as a function of displacement magnitudes.
The idea of making $L_n$ depend on displacement magnitudes is to average over more CSI features when the UE moves slowly, and over fewer features when it moves quickly.
Thus, when we compute averaged feature vectors with a displacement-dependent (abbreviated as ``disp.-dep.'') window length, we compute $L_n$ in \fref{eq:moving_average} as
\begin{align}
    L_n = \left\lceil \frac{a}{\|\sum_{k=n-10}^{n+10}\tilde{\bmdelta}_{k}\| + \epsilon} \right\rceil\!,
\end{align}
where $a, \epsilon > 0$ are user-defined parameters and $\lceil\cdot\rceil$ denotes rounding towards infinity. 
We set $\tilde{\bmdelta}_{n+10} = \mathbf{0}_2$ and $\tilde{\bmdelta}_{n-10} = \mathbf{0}_2$ for $n+10>N$ and $n-10<0$, respectively.

After averaging, the CSI features $\{\tilde{\bmf}_n\}_{n=0}^N$ are grouped into triangle sets as discussed in \fref{sec:ML_pipeline}.
The UE positioning performance results in \fref{sec:results} for the \WiFi scenarios were obtained from two variants of averaged feature vectors: (i) averaged with a fixed window length $L = 100$ and (ii) averaged with a disp.-dep. window size $L_n$.
\fref{sec:impact_of_L} presents UE positioning performance results for several fixed values $L$ in comparison with a disp.-dep. window size $L_n$.

\section{Baseline Methods}
\label{sec:baselines}

We now introduce three baselines that we use for comparison: (i) a CSI-based neural positioning method that utilizes external UE reference positions for training, (ii) a CC-based approach, which also utilizes estimated UE displacements and TDoA information, and (iii) a CSI fingerprinting approach with pseudo reference positions based on IMU data.

\subsection{Baseline 1 (External)}
\label{sec:baseline_1_external}
This baseline performs CSI-based neural positioning that uses external reference positions for training. 
We use the same MLP as in \fref{sec:ML_pipeline} and train it with the same CSI features, using a conventional mean-squared error (MSE) loss between the model output and the ground-truth UE positions.
This baseline serves as a performance reference to asses what could be achieved if external supervision is available.

\subsection{Baseline 2 (Internal)}
\label{sec:baseline_2_Internal}
This baseline emulates the method proposed in~\cite{ahadi2025displacementCC}; a CC approach that uses two reference inputs: (i) UE displacement magnitudes and (ii) TDoA measurements.
The following describes this baseline in detail, closely following~\cite{ahadi2025displacementCC}.

We compute UE displacement magnitudes $\{l_m\}_{m=0}^{N-\bar{V}}$ from our estimated displacements $\{\tilde{\bmdelta}_{n}\}_{n=0}^{N-1}$ as
\begin{equation}
    l_m = \big\lVert \textstyle \sum_{n=m}^{m+\bar{V}-1} \tilde{\bmdelta}_n \big\rVert,
\end{equation}
where we set $\bar{V} = 200$ across all experiments involving this baseline. Note the analogy to the leap increment parameter introduced in our proposed triangle method.

Since we do not have TDoA measurements in our datasets, we model TDoA measurements as random variables drawn from a simulated Gaussian distribution given our recorded reference UE positions and estimates of our AP positions.
We model a Gaussian randomness for our TDoA measurements to account for uncertainties in real-world TDoA estimation.
For each scenario, we declare one AP as the TDoA reference AP and denote its estimated position as $\bmx_{\text{ref}}$.
First, we compute the means in our simulated Gaussian TDoA distributions per time $m$ and AP $i$ as
\begin{equation}
    \tau_{m,i} = \frac{1}{c} \big( \lVert \bmx_{\text{ref}} - \bmx_m \rVert - \lVert \bmx_{\text{AP},i} - \bmx_m \rVert \big),
\end{equation}
for $m = 0, \ldots, N-\bar{V}$, where $c$ denotes the speed of light, $\bmx_m$ the recorded reference UE position at time $m$, and $\bmx_{\text{AP},i}$ the estimated position of AP $i$.
Then, we sample TDoA measurements as follows: $\tilde{\tau}_{m,i} \sim \normal(\tau_{m,i}, 3\ns^2)$. 
We choose a variance of $3\ns^2$ based on the estimation error magnitudes reported in~\cite{Schauer2013TDoA,Ma2022TDoA}.

The loss function for machine-learning training of this baseline, in line with the one proposed by \cite{ahadi2025displacementCC}, is given as
\be
\mathfrak{L}_2(\bmtheta) = \mathfrak{L}_{2d}(\bmtheta) + \mathfrak{L}_{2\tau}(\bmtheta)
\ee
with 
\begin{equation}
\mathfrak{L}_{2d}(\bmtheta)
= \sum_{m=0}^{N-\bar{V}} \Big| \| \bmg_{\bmtheta}(\bmf_{m+\bar{V}}) - \bmg_{\bmtheta}(\bmf_{m})\|  - l_m \Big|
\end{equation}
and
\begin{align}
\mathfrak{L}_{2\tau}(\bmtheta)
= \, & \sum_{m=0}^{N-\bar{V}} \sum_{i \in \setR}  \,  \Big| \| \bmx_{\text{ref}} - \bmg_{\bmtheta}(\bmf_{m})  \|  \, - \nonumber \\
& \qquad \qquad \,\,\, \|   \bmx_{\text{AP},i} - \bmg_{\bmtheta}(\bmf_{m})  \| - c\tilde{\tau}_{m,i} \Big| ,
\end{align}
where $\setR$ is the set of APs.

\subsection{Baseline 3 (Internal)}
\label{sec:baseline_3_Internal}
This baseline emulates the method proposed in~\cite{Ermolov2023imupos}, which is a CSI fingerprinting approach with IMU data-derived reference positions.
The method follows a two-stage procedure: first, absolute positions are obtained through IMU integration; second, a machine-learning model is trained analogously to the baseline in \fref{sec:baseline_1_external}, using these estimated absolute positions as reference.

We emulate this approach as follows. 
Recall the displacement loss in \fref{eq:max_likelihood} and the anchor loss in \fref{eq:negativ_log_likelihood_anchor}. As mentioned in \fref{sec:operatingprinciple}, one can estimate
the set of $N+1$ UE positions by minimizing the sum of these two losses.
First, we set $2\tilde{E}_n = 1$ in \fref{eq:max_likelihood} and $2\tilde{E}_a = 1$ in \fref{eq:negativ_log_likelihood_anchor}.
Second, we estimate the UE positions $\{\tilde{\bmx}_n\}_{n=0}^N$ by computing the closed-form solution of this least-squares problem.
Third, we use the same MLP as in \fref{sec:ML_pipeline} and train it with the same CSI features, using a conventional MSE loss between the model output and the estimated UE positions $\{\tilde{\bmx}_n\}_{n=0}^N$, analogous to the baseline in \fref{sec:baseline_1_external}.

\begin{figure*}[t!]
  \centering
  \subfloat[GT positions: \WiFi Small Meeting Room ]{
    \includegraphics[width=0.33\textwidth]{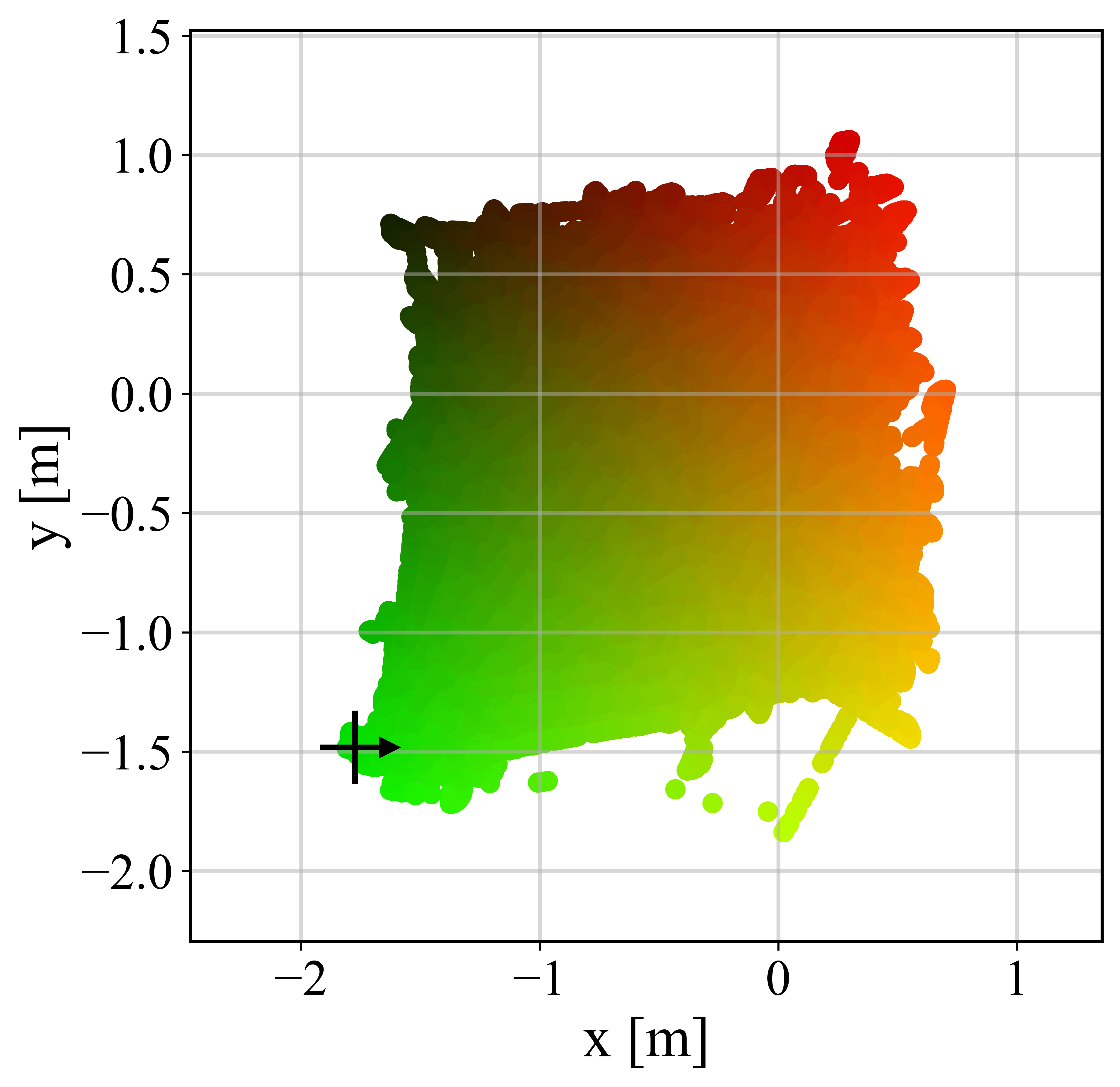}%
    \label{fig:GT_Test_Map_WiFi_Small_Room}%
  }\hfill
  \subfloat[GT positions: \WiFi Large Meeting Room]{
    \includegraphics[width=0.32\textwidth]{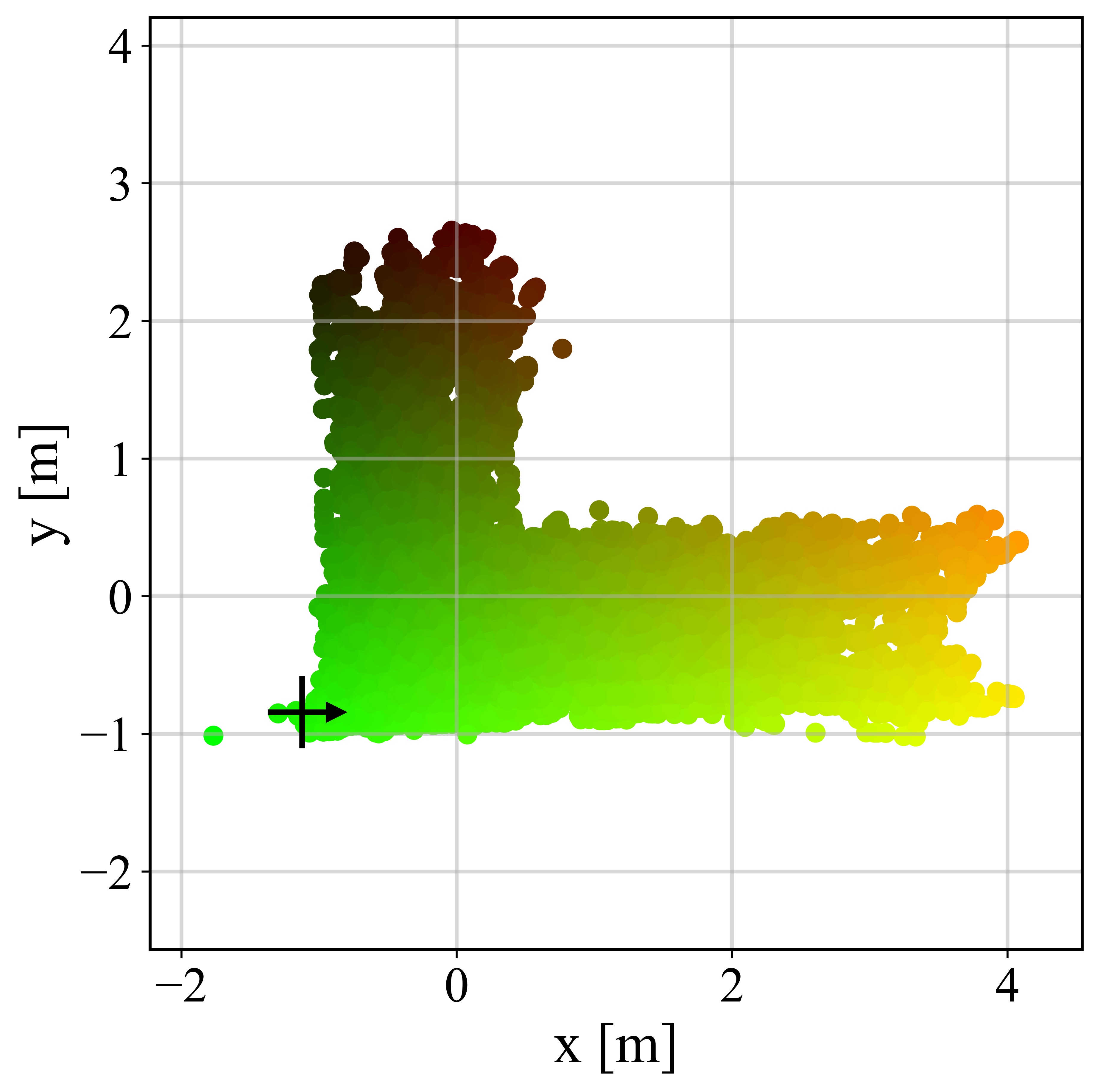}%
    \label{fig:GT_Test_Map_WiFi_Large_Room}%
  }\hfill
  \subfloat[GT positions: 5G Office]{
    \includegraphics[width=0.32\textwidth]{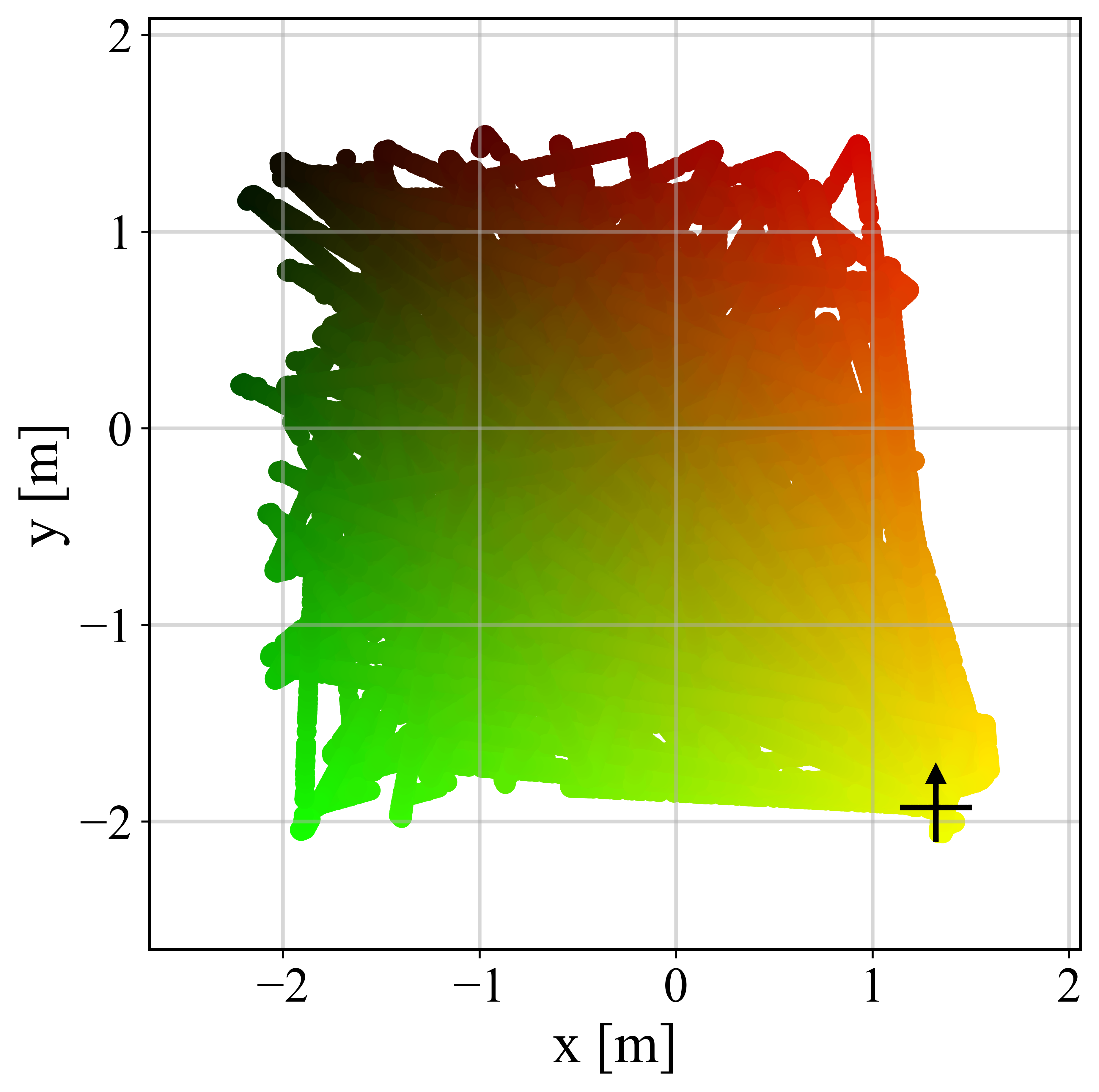}%
    \label{fig:GT_Test_Map_5G}%
  }

  \vspace{0.5em} 

  \subfloat[Estimator Output: \WiFi Small Meeting Room]{
    \includegraphics[width=0.33\textwidth]{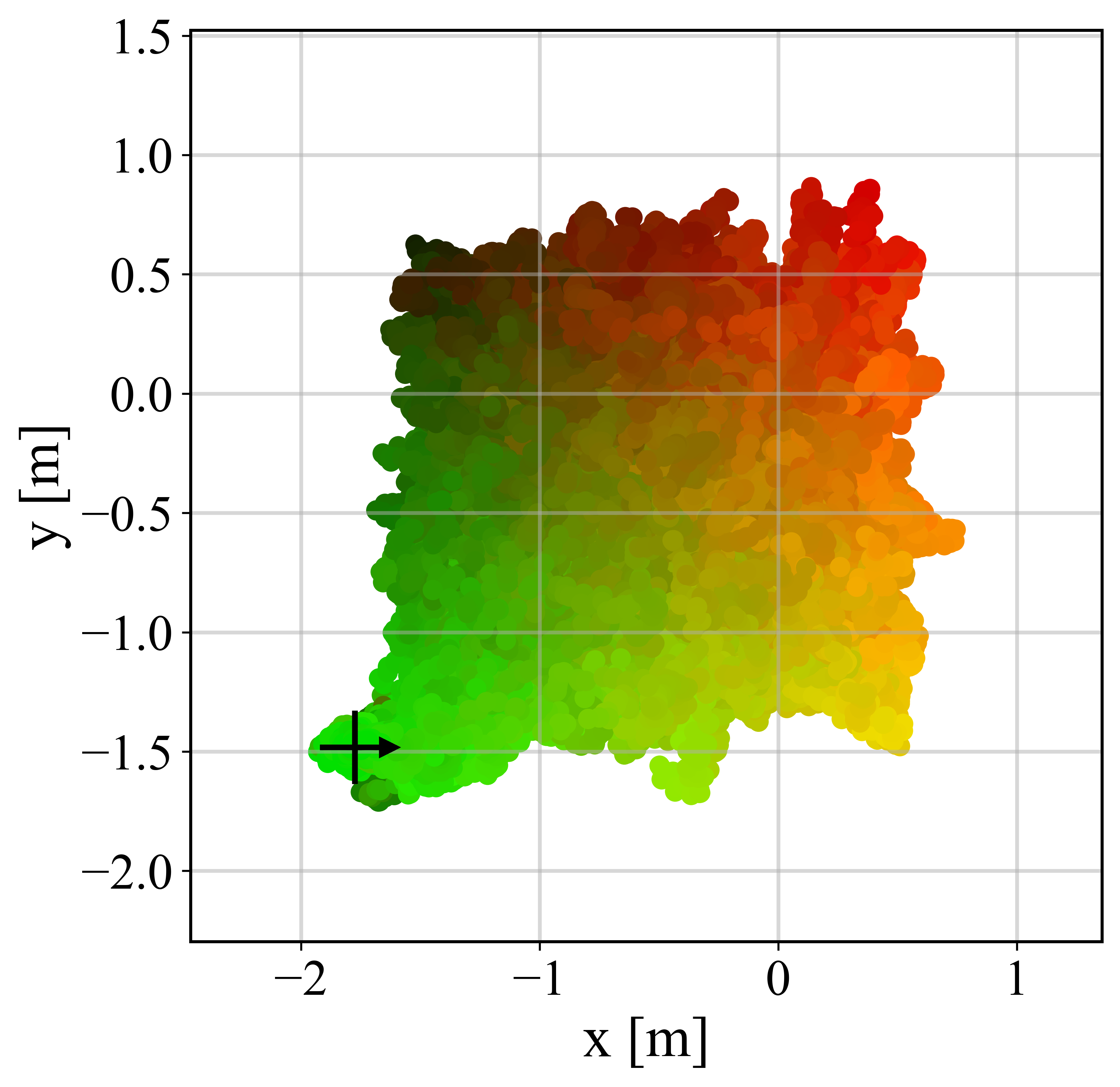}%
    \label{fig:Triangle_Error_Map_WiFi_Small_Room}%
  }\hfill
  \subfloat[Estimator Output: \WiFi Large Meeting Room]{
    \includegraphics[width=0.32\textwidth]{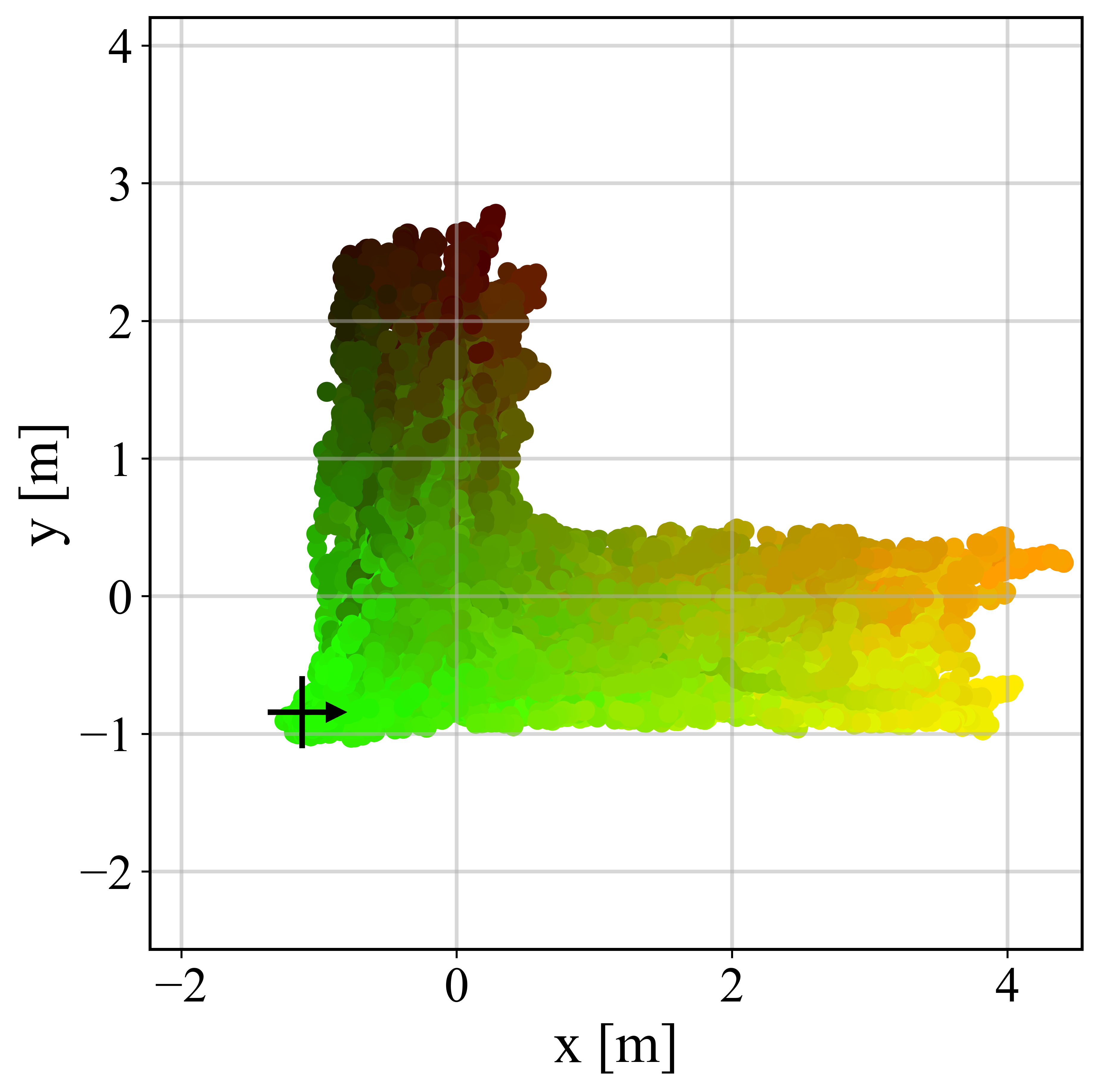}%
    \label{fig:Triangle_Error_Map_WiFi_Large_Room}%
  }\hfill
  \subfloat[Estimator Output: 5G Office]{
    \includegraphics[width=0.32\textwidth]{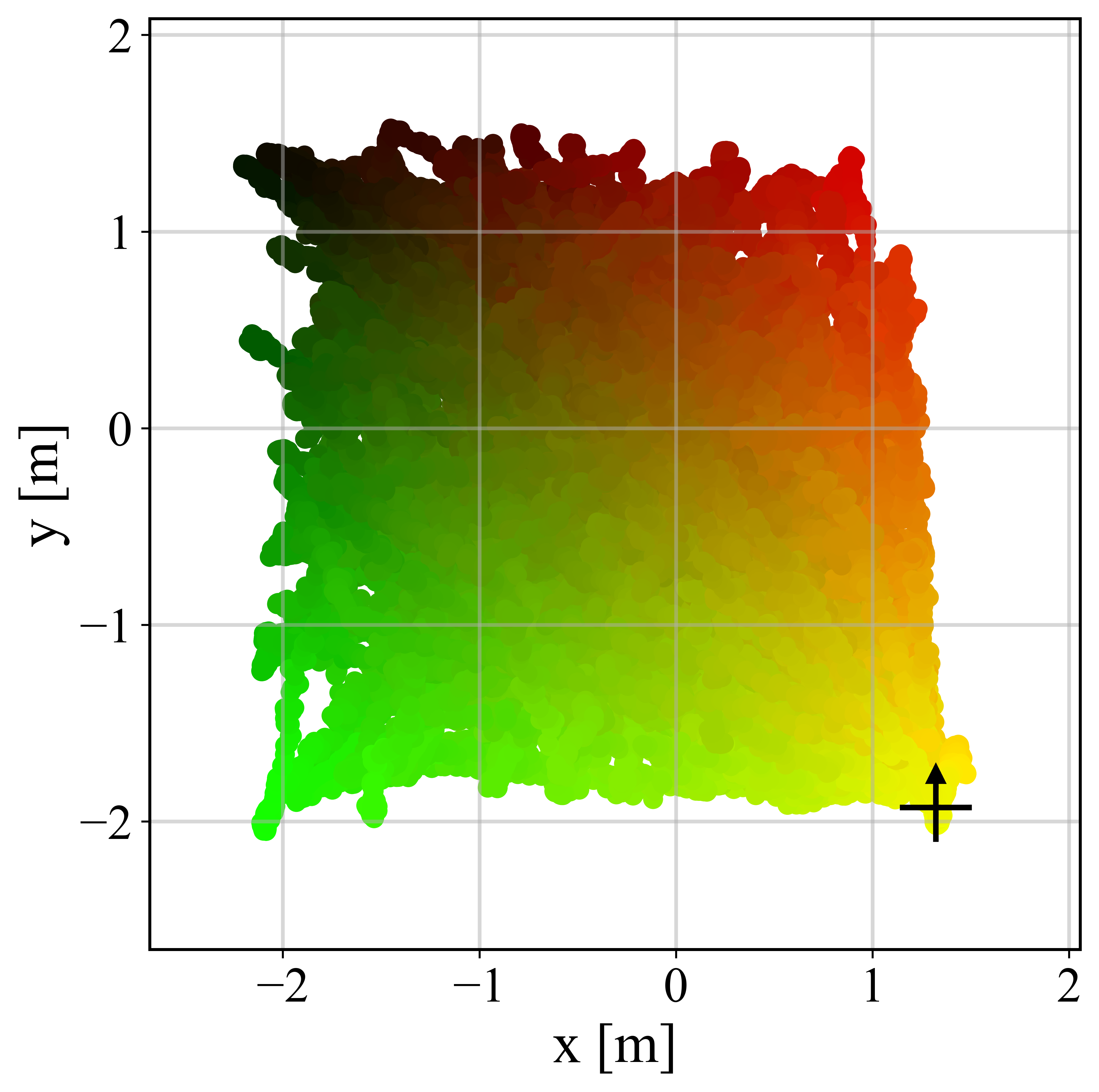}%
    \label{fig:Triangle_Error_Map_5G}%
  }

\vspace{0.1cm}

  \caption{Test set ground-truth (GT) positions (top row) and estimator outputs (bottom row) across the three measured scenarios. (a) and (d) correspond to the \WiFi small meeting room scenario, (b) and (e) to the the \WiFi large meeting room scenario, and (c) and (f) to the 5G office scenario with human activity. In all plots, the anchor position and its orientation are indicated by a black cross with an arrow. The arrow denotes the initial orientation of the UE platform.}
  \label{fig:Test_Maps_and_Error_Maps}
\end{figure*}

\begin{figure*}[t]
  \centering
  \subfloat[\WiFi Small Meeting Room]{
    \includegraphics[width=0.32\textwidth]{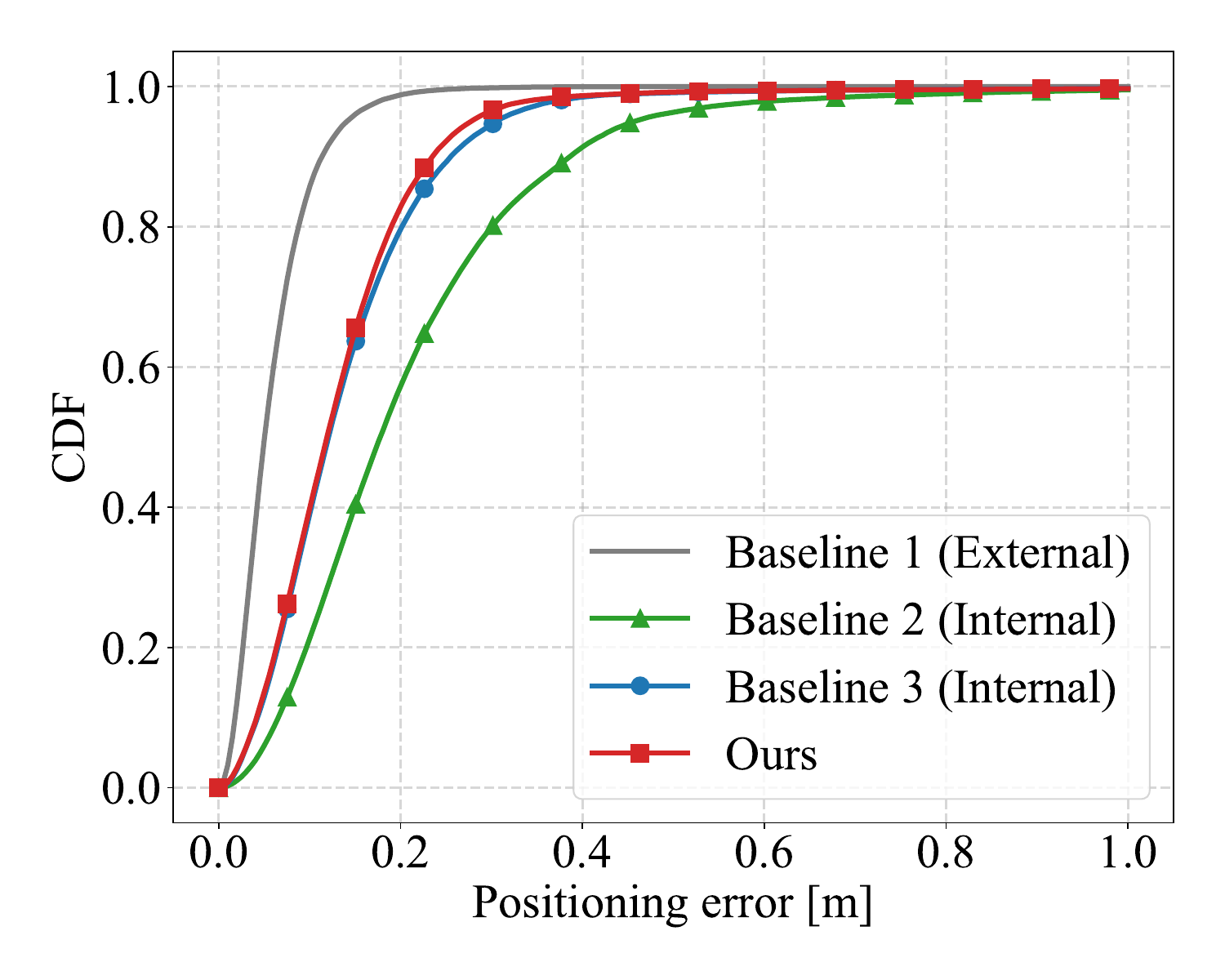}%
    \label{fig:CDF_WiFi_Small_MR}%
  }\hfill
  \subfloat[\WiFi Large Meeting Room]{
    \includegraphics[width=0.32\textwidth]{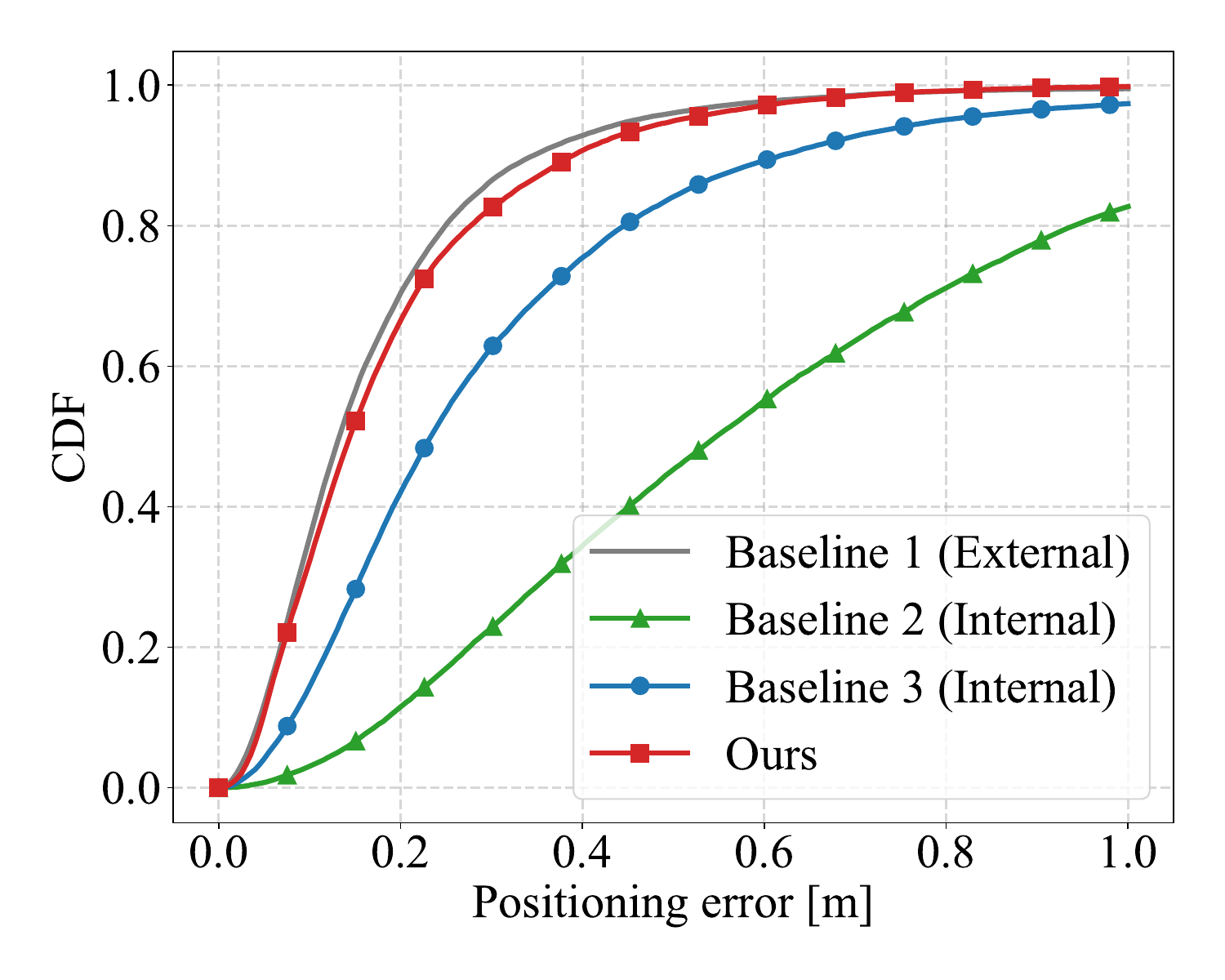}%
    \label{fig:CDF_WiFi_Large_MR}%
  }\hfill
  \subfloat[5G Office]{
    \includegraphics[width=0.32\textwidth]{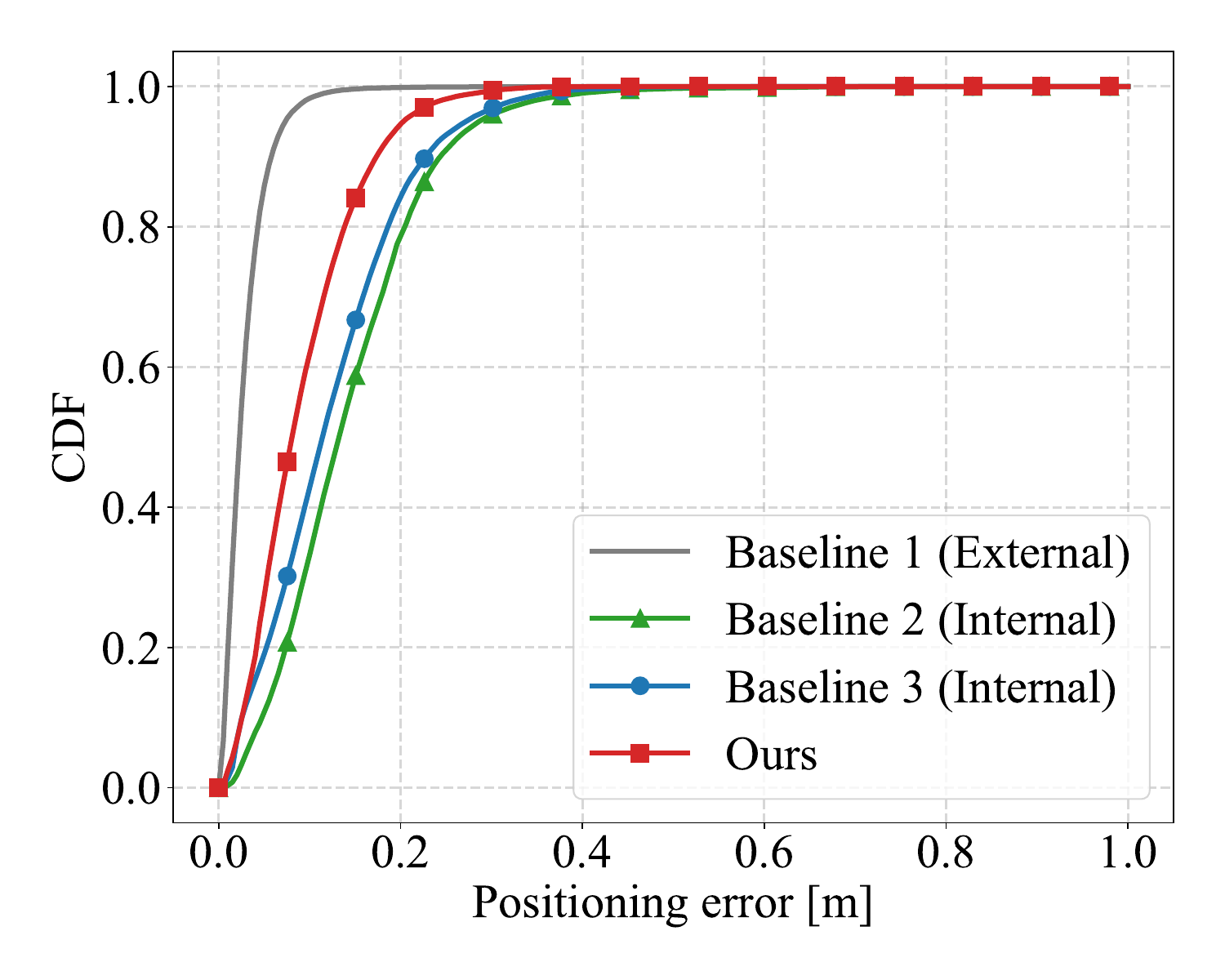}%
    \label{fig:5G_Office}%
  }
  
  \vspace{0.1cm}

  \caption{CDF of positioning errors for different methods, including Baseline 1 using external ground-truth position labels, Baseline 2, Baseline 3, and the proposed approach. For the \WiFi datasets, we used disp.-dep. features. Results are shown for (a) the \WiFi small meeting room dataset, (b) the \WiFi large meeting room dataset, and (c) the 5G office dataset.}
  \label{fig:CDF_plot}
\end{figure*}

\section{Experimental Results}
\label{sec:results}

We now show experimental UE positioning performance results using our neural positioning pipeline proposed in \fref{sec:ourmethod} as well as the baselines from~\fref{sec:baselines} for the three scenarios detailed in \fref{sec:measured_scenarios}.
We also investigate the impacts of (i) the size of the leap increments~$V$ in the triangle construction and (ii) the \WiFi data-specific time-averaging window~$L_n$.

 \subsection{Performance Metrics}

We evaluate UE positioning performance by measuring the absolute positioning error, which we define as the Euclidean distance between the positioning function output and the recorded ground-truth UE position with CSI measurements from a test-set that was excluded during training. 
We report the mean, median, and $95$th percentile positioning error, and also provide empirical cumulative distribution functions (CDFs) of the absolute positioning errors. 

\begin{table}[t]
\centering
\caption{Absolute positioning error [\cm].}
\label{tbl:abs_error}
\begin{tabular}{@{}p{4.6cm}ccc@{}}
\toprule
\textbf{Method} & \textbf{Mean} & \textbf{Median} & \textbf{$95$th\,\%} \\ \midrule
\multicolumn{4}{c}{\textit{\WiFi Small Meeting Room }} \\ \midrule
Baseline 1 (External) (avg. $L = 100$)         & 9.9 & 7.8 & 24.0 \\
Baseline 1 (External) (avg. disp.-dep.)    & 6.0 & 5.0 & 14.1 \\
Baseline 2 (Internal) (avg. disp.-dep.)   & 21.0 & 17.8 & 45.6 \\
Baseline 3 (Internal) (avg. disp.-dep.)   & 14.2 & 12.1 & 30.5 \\
Ours (avg. $L = 100$)        & 16.5 & 14.9 & 33.6 \\
\textbf{Ours (avg. disp.-dep.)}   & \textbf{13.5} & \textbf{11.9} & \textbf{27.7} \\ \midrule
\multicolumn{4}{c}{\textit{\WiFi Large Meeting Room }} \\ \midrule
Baseline 1 (External) (avg. $L = 100$)         & 18.7 & 13.8 & 50.0 \\
Baseline 1 (External) (avg. disp.-dep.)    & 17.6 & 13.2 & 45.6 \\
Baseline 2 (Internal) (avg. disp.-dep.) & 32.8 & 26.2 & 78.7 \\
Baseline 3 (Internal) (avg. disp.-dep.)   & 31.1 & 23.3 & 79.4 \\
Ours (avg. $L = 100$)        & 21.2 & 16.6 & 54.1 \\
\textbf{Ours (avg. disp.-dep.)}   & \textbf{18.9} & \textbf{14.5} & \textbf{50.2} \\ \midrule

\multicolumn{4}{c}{\textit{5G Office}} \\ \midrule
Baseline 1 (External)  &2.9 & 2.3 & 7.3 \\
Baseline 2 (Internal) & 14.3 & 13.2 & 28.6 \\
Baseline 3 (Internal)  & 12.4 & 11.4 & 27.1 \\
\textbf{Ours}        & \textbf{9.2} & \textbf{8.1} & \textbf{20.3} \\  \midrule
\end{tabular}
\end{table}

\subsection{Positioning Results}

\fref{fig:Test_Maps_and_Error_Maps} shows the ground-truth positions (top row) and estimated positions of our neural positioning pipeline (bottom row) for the three scenarios.
The color gradients are used to assist a visual comparison. 
We see that the UE positions predicted by our method closely match the ground-truth positions in all three scenarios, with only few outliers. 

\fref{tbl:abs_error} shows UE positioning error statistics (mean absolute error, median absolute error, and $95$th percentile absolute error) for our proposed neural positioning pipeline as well as the considered baseline methods. 
We see that our method achieves a positioning accuracy that is only slightly lower than that of the supervised positioning baseline (Baseline 1), which requires an external reference positioning system. In comparison with the two baselines that do not require external reference position information for training (Baselines 2 and 3), our approach consistently outperforms those methods in all three metrics. 
We see that for the two \WiFi-based scenarios, Small Meeting Room and Large Meeting Room, our method achieves a median absolute error of $11.9$\,cm and $14.5$\,cm, respectively, and only $8.1$\,cm median absolute error for the 5G-based Office scenario. 
The 5G-based scenario performs better as (i) it builds upon high-quality COTS O-RU hardware, (ii) utilizes a wider bandwidth, (iii) operates at a lower carrier frequency, and (iv) performs learning from a much denser set of CSI measurements (both in time and frequency) compared to the \WiFi-based scenario.

\fref{fig:CDF_plot} shows empirical CDFs of the positioning error for the three scenarios and the three baselines. 
We see that our approach consistently outperforms the two baselines that do not require external reference positions for training (Baselines~2 and~3) and approaches the accuracy of the supervised positioning baseline (Baseline 1), which requires an external reference.

These experiments demonstrate that centimeter-level neural positioning is possible in various scenarios with COTS hardware (for the UE, APs, as well as the moving robot platform) without the need for an external reference positioning system. This observation implies that our method enables one to train (and retrain) robust neural positioning functions over very large areas and in complex scenarios in an inexpensive manner. 

\begin{rem}
All of these UE positioning results are single-shot, i.e., take one (or, for the \WiFi scenario, multiple) CSI features and generate \emph{one} position estimate. Methods that track UE position over time, e.g., using Kalman filters, are expected to significantly reduce outliers and achieve better positioning performance. Such methods are, however, not pursued further. 
\end{rem}

\subsection{Impacts of Leap Increment and Window Length}
\subsubsection{Impact of Leap Increment $V$}
\label{sec:leap_increment_effect}

\begin{figure}[tp]
  \centering
  \includegraphics[width=\linewidth]{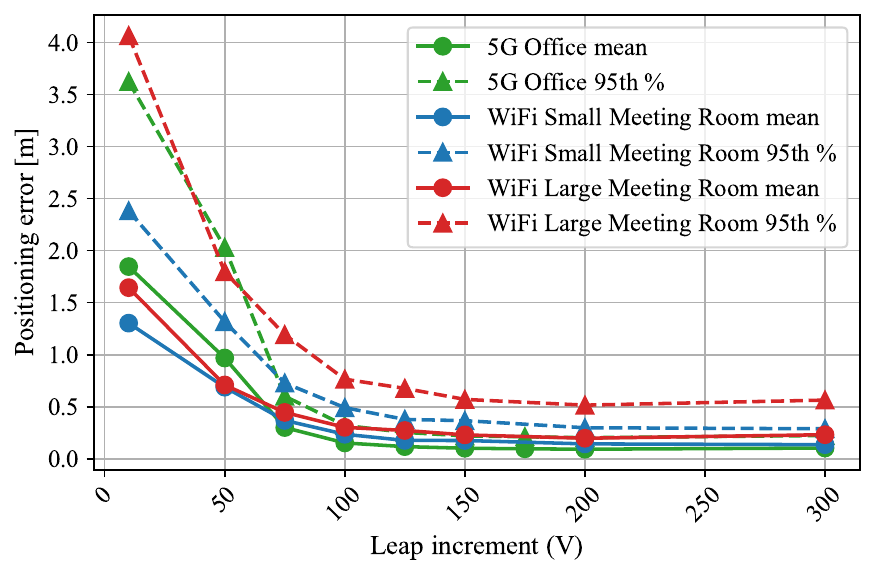}
  \vspace{-0.5cm}
  \caption{Impact of the leap increment parameter $V$: mean and $95$th percentile positioning errors using our proposed method across all three scenarios. Leap increments of $100$ or more result in the lowest positioning errors.}
  \label{fig:error_leap_increment}
\end{figure}

\fref{fig:error_leap_increment} shows the mean and $95$th  percentile errors for leap increment parameter $V$ ranging from $10$ to $300$ in all three scenarios.
We see that increasing the leap increment parameter $V$ quickly (and significantly) improves positioning accuracy, saturating after about $V=100$. 

We note that the CSI feature sampling intervals differ for the three datasets. As a result, a leap increment of $V = 100$ corresponds to sampling intervals of approximately $2\,$s, $5\,$s, and $10\,$s for the 5G Office, the \WiFi Small Meeting Room, and the \WiFi Large Meeting Room datasets, respectively. 
Consequently, since the robot platform moved at a constant velocity, the average triangle side lengths observed were $29\,$cm, $62\,$cm, and $20\,$cm for the \WiFi small meeting room, \WiFi large meeting room, and 5G office scenarios, respectively.
The saturating behavior for large leap increments (e.g., $V>100$) can be explained by the fact that a sufficiently large leap increment is required for the CSI features to properly capture large-scale fading effects of the wireless channel.

\begin{figure}[tp]
  \centering
  \includegraphics[width=\linewidth]{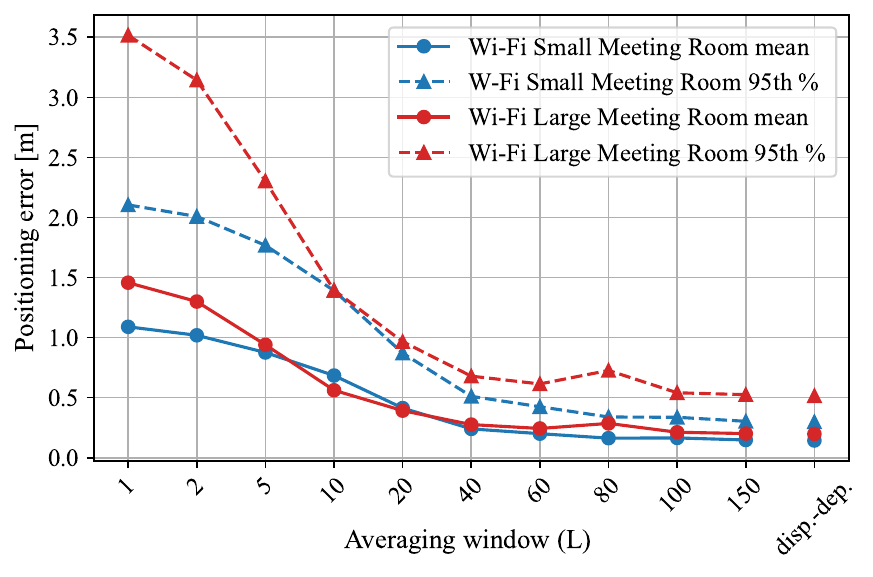}
  \vspace{-0.5cm}
  \caption{Impact of the \WiFi dataset-specific time-averaging window length~$L$: mean and $95$th  percentile positioning errors using our proposed method across the two \WiFi scenarios.}
  \label{fig:Average_Window}
\end{figure}

\subsubsection{Impact of Window Length $L$}
\label{sec:impact_of_L}
\fref{fig:Average_Window} shows the mean and $95$th percentile positioning errors for time-averaging window sizes $L$ described in \fref{sec:wifi_feature_averaging}, ranging from~$1$ to $150$ for the two \WiFi datasets (Small Meeting Room and Large Meeting Room). 
We see that increasing the window length $L$ quickly (and significantly) reduces the positioning error and saturates after around $L=40$.
\fref{fig:Average_Window} also shows the displacement-dependent window size (indicated by ``disp.-dep.''), which consistently achieves the best overall positioning performance as can also be observed in \fref{tbl:abs_error}.

\section{Conclusions and Future Work}
\label{sec:conclusions}

We have proposed a neural positioning pipeline that trains functions which accurately map measured CSI to position without the need for an external reference positioning system. 
Our training method utilizes relative displacement measurements, which could, for example, be displacement commands executed on a robot platform that passes through the area of interest. 
We have validated the effectiveness of our approach using three real-world CSI datasets measured with IEEE 802.11 \WiFi and 5G NR wireless systems and using COTS hardware for the UE, the APs, and the mobile robot platform. 
Experiments in three different scenarios demonstrate that our neural positioning pipeline consistently achieves centimeter-level accuracy and approaches the performance of state-of-the-art CSI-based neural positioning systems that are trained with ground-truth labels from an external reference positioning system. 
These results imply that our proposed method enables one to train (and retrain) robust neural positioning functions over very large
areas and in complex scenarios in an inexpensive manner.

There exist a range of possibilities for future work. 
First, using more accurate displacement error models is expected to further improve the performance of our proposed neural positioning pipeline. 
Second, developing methods that smoothen UE position over time, e.g., using Kalman filters or more advanced machine-learning models, are expected to further reduce outliers and improve positioning accuracy.
Third, conducting large-scale experiments covering areas of entire building floors, would further confirm the effectiveness of our method.

\begin{rem}
We will make the three CSI datasets discussed in \fref{tbl:datasets} as well as the code used to perform our experiments available to the public after the peer-review process.
\end{rem}

\balance
\bibliographystyle{IEEEtran}
\bibliography{bib/VIPabbrv,bib/confs-jrnls,bib/publishers,bib/my_bib}

\end{document}

%% file: macros/vmr-symbols-vecbold.tex
\usepackage{amssymb}
\usepackage{amsfonts}
\usepackage{mathrsfs}
\usepackage{xspace}
\usepackage{bm}
\usepackage{upgreek}

\newcommand{\safemath}[2]{\newcommand{#1}{\ensuremath{#2}\xspace}}

\safemath{\bma}{\mathbf{a}}
\safemath{\bmb}{\mathbf{b}}
\safemath{\bmc}{\mathbf{c}}
\safemath{\bmd}{\mathbf{d}}
\safemath{\bme}{\mathbf{e}}
\safemath{\bmf}{\mathbf{f}}
\safemath{\bmg}{\mathbf{g}}
\safemath{\bmh}{\mathbf{h}}
\safemath{\bmi}{\mathbf{i}}
\safemath{\bmj}{\mathbf{j}}
\safemath{\bmk}{\mathbf{k}}
\safemath{\bml}{\mathbf{l}}
\safemath{\bmm}{\mathbf{m}}
\safemath{\bmn}{\mathbf{n}}
\safemath{\bmo}{\mathbf{o}}
\safemath{\bmp}{\mathbf{p}}
\safemath{\bmq}{\mathbf{q}}
\safemath{\bmr}{\mathbf{r}}
\safemath{\bms}{\mathbf{s}}
\safemath{\bmt}{\mathbf{t}}
\safemath{\bmu}{\mathbf{u}}
\safemath{\bmv}{\mathbf{v}}
\safemath{\bmw}{\mathbf{w}}
\safemath{\bmx}{\mathbf{x}}
\safemath{\bmy}{\mathbf{y}}
\safemath{\bmz}{\mathbf{z}}
\safemath{\bmzero}{\mathbf{0}}
\safemath{\bmone}{\mathbf{1}}

\bmdefine{\biad}{a}
\bmdefine{\bibd}{b}
\bmdefine{\bicd}{c}
\bmdefine{\bidd}{d}
\bmdefine{\bied}{e}
\bmdefine{\bifd}{f}
\bmdefine{\bigd}{g}
\bmdefine{\bihd}{h}
\bmdefine{\biid}{i}
\bmdefine{\bijd}{j}
\bmdefine{\bikd}{k}
\bmdefine{\bild}{l}
\bmdefine{\bimd}{m}
\bmdefine{\bind}{n}
\bmdefine{\biod}{o}
\bmdefine{\bipd}{p}
\bmdefine{\biqd}{q}
\bmdefine{\bird}{r}
\bmdefine{\bisd}{s}
\bmdefine{\bitd}{t}
\bmdefine{\biud}{u}
\bmdefine{\bivd}{v}
\bmdefine{\biwd}{w}
\bmdefine{\bixd}{x}
\bmdefine{\biyd}{y}
\bmdefine{\bizd}{z}

\bmdefine{\bixid}{\xi}
\bmdefine{\bilambdad}{\lambda}
\bmdefine{\bimud}{\mu}
\bmdefine{\bithetad}{\theta}
\bmdefine{\biphid}{\phi}
\bmdefine{\bideltad}{\delta}

\safemath{\bmia}{\biad}
\safemath{\bmib}{\bibd}
\safemath{\bmic}{\bicd}
\safemath{\bmid}{\bidd}
\safemath{\bmie}{\bied}
\safemath{\bmif}{\bifd}
\safemath{\bmig}{\bigd}
\safemath{\bmih}{\bihd}
\safemath{\bmii}{\biid}
\safemath{\bmij}{\bijd}
\safemath{\bmik}{\bikd}
\safemath{\bmil}{\bild}
\safemath{\bmim}{\bimd}
\safemath{\bmin}{\bind}
\safemath{\bmio}{\biod}
\safemath{\bmip}{\bipd}
\safemath{\bmiq}{\biqd}
\safemath{\bmir}{\bird}
\safemath{\bmis}{\bisd}
\safemath{\bmit}{\bitd}
\safemath{\bmiu}{\biud}
\safemath{\bmiv}{\bivd}
\safemath{\bmiw}{\biwd}
\safemath{\bmix}{\bixd}
\safemath{\bmiy}{\biyd}
\safemath{\bmiz}{\bizd}

\safemath{\bmxi}{\bixid}
\safemath{\bmlambda}{\bilambdad}
\safemath{\bmmu}{\bimud}
\safemath{\bmtheta}{\bithetad}
\safemath{\bmphi}{\biphid}
\safemath{\bmdelta}{\bideltad}

\safemath{\bA}{\mathbf{A}}
\safemath{\bB}{\mathbf{B}}
\safemath{\bC}{\mathbf{C}}
\safemath{\bD}{\mathbf{D}}
\safemath{\bE}{\mathbf{E}}
\safemath{\bF}{\mathbf{F}}
\safemath{\bG}{\mathbf{G}}
\safemath{\bH}{\mathbf{H}}
\safemath{\bI}{\mathbf{I}}
\safemath{\bJ}{\mathbf{J}}
\safemath{\bK}{\mathbf{K}}
\safemath{\bL}{\mathbf{L}}
\safemath{\bM}{\mathbf{M}}
\safemath{\bN}{\mathbf{N}}
\safemath{\bO}{\mathbf{O}}
\safemath{\bP}{\mathbf{P}}
\safemath{\bQ}{\mathbf{Q}}
\safemath{\bR}{\mathbf{R}}
\safemath{\bS}{\mathbf{S}}
\safemath{\bT}{\mathbf{T}}
\safemath{\bU}{\mathbf{U}}
\safemath{\bV}{\mathbf{V}}
\safemath{\bW}{\mathbf{W}}
\safemath{\bX}{\mathbf{X}}
\safemath{\bY}{\mathbf{Y}}
\safemath{\bZ}{\mathbf{Z}}

\safemath{\bZero}{\mathbf{0}}
\safemath{\bOne}{\mathbf{1}}
\safemath{\bDelta}{\mathbf{\Delta}}
\safemath{\bLambda}{\mathbf{\UpLambda}}
\safemath{\bPhi}{\mathbf{\Upphi}}
\safemath{\bSigma}{\mathbf{\Upsigma}}
\safemath{\bOmega}{\mathbf{\Upomega}}
\safemath{\bTheta}{\mathbf{\Uptheta}}

\bmdefine{\biAd}{A}
\bmdefine{\biBd}{B}
\bmdefine{\biCd}{C}
\bmdefine{\biDd}{D}
\bmdefine{\biEd}{E}
\bmdefine{\biFd}{F}
\bmdefine{\biGd}{G}
\bmdefine{\biHd}{H}
\bmdefine{\biId}{I}
\bmdefine{\biJd}{J}
\bmdefine{\biKd}{K}
\bmdefine{\biLd}{L}
\bmdefine{\biMd}{M}
\bmdefine{\biOd}{N}
\bmdefine{\biPd}{O}
\bmdefine{\biQd}{P}
\bmdefine{\biRd}{R}
\bmdefine{\biSd}{S}
\bmdefine{\biTd}{T}
\bmdefine{\biUd}{U}
\bmdefine{\biVd}{V}
\bmdefine{\biWd}{W}
\bmdefine{\biXd}{X}
\bmdefine{\biYd}{Y}
\bmdefine{\biZd}{Z}

\bmdefine{\biDelta}{\Delta}
\bmdefine{\biLambda}{\Lambda}
\bmdefine{\biPhi}{\Phi}
\bmdefine{\biSigma}{\Sigma}
\bmdefine{\biOmega}{\Omega}
\bmdefine{\biTheta}{\Theta}

\safemath{\bimA}{\biAd}
\safemath{\bimB}{\biBd}
\safemath{\bimC}{\biCd}
\safemath{\bimD}{\biDd}
\safemath{\bimE}{\biEd}
\safemath{\bimF}{\biFd}
\safemath{\bimG}{\biGd}
\safemath{\bimH}{\biHd}
\safemath{\bimI}{\biId}
\safemath{\bimJ}{\biJd}
\safemath{\bimK}{\biKd}
\safemath{\bimL}{\biLd}
\safemath{\bimM}{\biMd}
\safemath{\bimN}{\biNd}
\safemath{\bimO}{\biOd}
\safemath{\bimP}{\biPd}
\safemath{\bimQ}{\biQd}
\safemath{\bimR}{\biRd}
\safemath{\bimS}{\biSd}
\safemath{\bimT}{\biTd}
\safemath{\bimU}{\biUd}
\safemath{\bimV}{\biVd}
\safemath{\bimW}{\biWd}
\safemath{\bimX}{\biXd}
\safemath{\bimY}{\biYd}
\safemath{\bimZ}{\biZd}

\safemath{\bimDelta}{\biDelta}
\safemath{\bimLambda}{\biLambda}
\safemath{\bimPhi}{\biPhi}
\safemath{\bimSigma}{\biSigma}
\safemath{\bimOmega}{\biOmega}
\safemath{\bimTheta}{\biTheta}

\safemath{\setA}{\mathcal{A}}
\safemath{\setB}{\mathcal{B}}
\safemath{\setC}{\mathcal{C}}
\safemath{\setD}{\mathcal{D}}
\safemath{\setE}{\mathcal{E}}
\safemath{\setF}{\mathcal{F}}
\safemath{\setG}{\mathcal{G}}
\safemath{\setH}{\mathcal{H}}
\safemath{\setI}{\mathcal{I}}
\safemath{\setJ}{\mathcal{J}}
\safemath{\setK}{\mathcal{K}}
\safemath{\setL}{\mathcal{L}}
\safemath{\setM}{\mathcal{M}}
\safemath{\setN}{\mathcal{N}}
\safemath{\setO}{\mathcal{O}}
\safemath{\setP}{\mathcal{P}}
\safemath{\setQ}{\mathcal{Q}}
\safemath{\setR}{\mathcal{R}}
\safemath{\setS}{\mathcal{S}}
\safemath{\setT}{\mathcal{T}}
\safemath{\setU}{\mathcal{U}}
\safemath{\setV}{\mathcal{V}}
\safemath{\setW}{\mathcal{W}}
\safemath{\setX}{\mathcal{X}}
\safemath{\setY}{\mathcal{Y}}
\safemath{\setZ}{\mathcal{Z}}
\safemath{\emptySet}{\varnothing}

\safemath{\colA}{\mathscr{A}}
\safemath{\colB}{\mathscr{B}}
\safemath{\colC}{\mathscr{C}}
\safemath{\colD}{\mathscr{D}}
\safemath{\colE}{\mathscr{E}}
\safemath{\colF}{\mathscr{F}}
\safemath{\colG}{\mathscr{G}}
\safemath{\colH}{\mathscr{H}}
\safemath{\colI}{\mathscr{I}}
\safemath{\colJ}{\mathscr{J}}
\safemath{\colK}{\mathscr{K}}
\safemath{\colL}{\mathscr{L}}
\safemath{\colM}{\mathscr{M}}
\safemath{\colN}{\mathscr{N}}
\safemath{\colO}{\mathscr{O}}
\safemath{\colP}{\mathscr{P}}
\safemath{\colQ}{\mathscr{Q}}
\safemath{\colR}{\mathscr{R}}
\safemath{\colS}{\mathscr{S}}
\safemath{\colT}{\mathscr{T}}
\safemath{\colU}{\mathscr{U}}
\safemath{\colV}{\mathscr{V}}
\safemath{\colW}{\mathscr{W}}
\safemath{\colX}{\mathscr{X}}
\safemath{\colY}{\mathscr{Y}}
\safemath{\colZ}{\mathscr{Z}}

\safemath{\opA}{\mathbb{A}}
\safemath{\opB}{\mathbb{B}}
\safemath{\opC}{\mathbb{C}}
\safemath{\opD}{\mathbb{D}}
\safemath{\opE}{\mathbb{E}}
\safemath{\opF}{\mathbb{F}}
\safemath{\opG}{\mathbb{G}}
\safemath{\opH}{\mathbb{H}}
\safemath{\opI}{\mathbb{I}}
\safemath{\opJ}{\mathbb{J}}
\safemath{\opK}{\mathbb{K}}
\safemath{\opL}{\mathbb{L}}
\safemath{\opM}{\mathbb{M}}
\safemath{\opN}{\mathbb{N}}
\safemath{\opO}{\mathbb{O}}
\safemath{\opP}{\mathbb{P}}
\safemath{\opQ}{\mathbb{Q}}
\safemath{\opR}{\mathbb{R}}
\safemath{\opS}{\mathbb{S}}
\safemath{\opT}{\mathbb{T}}
\safemath{\opU}{\mathbb{U}}
\safemath{\opV}{\mathbb{V}}
\safemath{\opW}{\mathbb{W}}
\safemath{\opX}{\mathbb{X}}
\safemath{\opY}{\mathbb{Y}}
\safemath{\opZ}{\mathbb{Z}}
\safemath{\opZero}{\mathbb{O}}
\safemath{\identityop}{\opI}

\safemath{\veca}{\bma}
\safemath{\vecb}{\bmb}
\safemath{\vecc}{\bmc}
\safemath{\vecd}{\bmd}
\safemath{\vece}{\bme}
\safemath{\vecf}{\bmf}
\safemath{\vecg}{\bmg}
\safemath{\vech}{\bmh}
\safemath{\veci}{\bmi}
\safemath{\vecj}{\bmj}
\safemath{\veck}{\bmk}
\safemath{\vecl}{\bml}
\safemath{\vecm}{\bmm}
\safemath{\vecn}{\bmn}
\safemath{\veco}{\bmo}
\safemath{\vecp}{\bmp}
\safemath{\vecq}{\bmq}
\safemath{\vecr}{\bmr}
\safemath{\vecs}{\bms}
\safemath{\vect}{\bmt}
\safemath{\vecu}{\bmu}
\safemath{\vecv}{\bmv}
\safemath{\vecw}{\bmw}
\safemath{\vecx}{\bmx}
\safemath{\vecy}{\bmy}
\safemath{\vecz}{\bmz}

\safemath{\veczero}{\bmzero}
\safemath{\vecone}{\bmone}
\safemath{\vecxi}{\bmxi}
\safemath{\veclambda}{\bmlambda}
\safemath{\vecmu}{\bmmu}
\safemath{\vectheta}{\bmtheta}
\safemath{\vecphi}{\bmphi}
\safemath{\vecdelta}{\bmdelta}

\safemath{\matA}{\bA}
\safemath{\matB}{\bB}
\safemath{\matC}{\bC}
\safemath{\matD}{\bD}
\safemath{\matE}{\bE}
\safemath{\matF}{\bF}
\safemath{\matG}{\bG}
\safemath{\matH}{\bH}
\safemath{\matI}{\bI}
\safemath{\matJ}{\bJ}
\safemath{\matK}{\bK}
\safemath{\matL}{\bL}
\safemath{\matM}{\bM}
\safemath{\matN}{\bN}
\safemath{\matO}{\bO}
\safemath{\matP}{\bP}
\safemath{\matQ}{\bQ}
\safemath{\matR}{\bR}
\safemath{\matS}{\bS}
\safemath{\matT}{\bT}
\safemath{\matU}{\bU}
\safemath{\matV}{\bV}
\safemath{\matW}{\bW}
\safemath{\matX}{\bX}
\safemath{\matY}{\bY}
\safemath{\matZ}{\bZ}
\safemath{\matzero}{\bmzero}

\safemath{\matDelta}{\bDelta}
\safemath{\matLambda}{\bLambda}
\safemath{\matPhi}{\bPhi}
\safemath{\matSigma}{\bSigma}
\safemath{\matOmega}{\bOmega}
\safemath{\matTheta}{\bTheta}

\safemath{\matidentity}{\matI}
\safemath{\matone}{\matO}

\safemath{\rnda}{A}
\safemath{\rndb}{B}
\safemath{\rndc}{C}
\safemath{\rndd}{D}
\safemath{\rnde}{E}
\safemath{\rndf}{F}
\safemath{\rndg}{G}
\safemath{\rndh}{H}
\safemath{\rndi}{I}
\safemath{\rndj}{J}
\safemath{\rndk}{K}
\safemath{\rndl}{L}
\safemath{\rndm}{M}
\safemath{\rndn}{N}
\safemath{\rndo}{O}
\safemath{\rndp}{P}
\safemath{\rndq}{Q}
\safemath{\rndr}{R}
\safemath{\rnds}{S}
\safemath{\rndt}{T}
\safemath{\rndu}{U}
\safemath{\rndv}{V}
\safemath{\rndw}{W}
\safemath{\rndx}{X}
\safemath{\rndy}{Y}
\safemath{\rndz}{Z}

\safemath{\rveca}{\bimA}
\safemath{\rvecb}{\bimB}
\safemath{\rvecc}{\bimC}
\safemath{\rvecd}{\bimD}
\safemath{\rvece}{\bimE}
\safemath{\rvecf}{\bimF}
\safemath{\rvecg}{\bimG}
\safemath{\rvech}{\bimH}
\safemath{\rveci}{\bimI}
\safemath{\rvecj}{\bimJ}
\safemath{\rveck}{\bimK}
\safemath{\rvecl}{\bimL}
\safemath{\rvecm}{\bimM}
\safemath{\rvecn}{\bimN}
\safemath{\rveco}{\bomO}
\safemath{\rvecp}{\bimP}
\safemath{\rvecq}{\bimQ}
\safemath{\rvecr}{\bimR}
\safemath{\rvecs}{\bimS}
\safemath{\rvect}{\bimT}
\safemath{\rvecu}{\bimU}
\safemath{\rvecv}{\bimV}
\safemath{\rvecw}{\bimW}
\safemath{\rvecx}{\bimX}
\safemath{\rvecy}{\bimY}
\safemath{\rvecz}{\bimZ}

\safemath{\rvecxi}{\bmxi}
\safemath{\rveclambda}{\bmlambda}
\safemath{\rvecmu}{\bmmu}
\safemath{\rvectheta}{\bmtheta}
\safemath{\rvecphi}{\bmphi}

\safemath{\rmatA}{\bimA}
\safemath{\rmatB}{\bimB}
\safemath{\rmatC}{\bimC}
\safemath{\rmatD}{\bimD}
\safemath{\rmatE}{\bimE}
\safemath{\rmatF}{\bimF}
\safemath{\rmatG}{\bimG}
\safemath{\rmatH}{\bimH}
\safemath{\rmatI}{\bimI}
\safemath{\rmatJ}{\bimJ}
\safemath{\rmatK}{\bimK}
\safemath{\rmatL}{\bimL}
\safemath{\rmatM}{\bimM}
\safemath{\rmatN}{\bimN}
\safemath{\rmatO}{\bimO}
\safemath{\rmatP}{\bimP}
\safemath{\rmatQ}{\bimQ}
\safemath{\rmatR}{\bimR}
\safemath{\rmatS}{\bimS}
\safemath{\rmatT}{\bimT}
\safemath{\rmatU}{\bimU}
\safemath{\rmatV}{\bimV}
\safemath{\rmatW}{\bimW}
\safemath{\rmatX}{\bimX}
\safemath{\rmatY}{\bimY}
\safemath{\rmatZ}{\bimZ}

\safemath{\rmatDelta}{\bimDelta}
\safemath{\rmatLambda}{\bimLambda}
\safemath{\rmatPhi}{\bimPhi}
\safemath{\rmatSigma}{\bimSigma}
\safemath{\rmatOmega}{\bimOmega}
\safemath{\rmatTheta}{\bimTheta}

%% file: macros/standard-macros.tex
\usepackage{amssymb}
\usepackage{amsfonts}
\usepackage{mathrsfs}
\usepackage{xspace}
\usepackage{bm}
\usepackage{fancyref}
\usepackage{textcomp}

\usepackage{multirow}
\usepackage{stmaryrd}

\newenvironment{textbmatrix}{	\setlength{\arraycolsep}{2.5pt}%
								\big[\begin{matrix}}{\end{matrix}\big]%
								\raisebox{0.08ex}{\vphantom{M}}}

\def\be{\begin{equation}}
\def\ee{\end{equation}}
\def\een{\nonumber \end{equation}}
\def\mat{\begin{bmatrix}}
\def\emat{\end{bmatrix}}
\def\btm{\begin{textbmatrix}}
\def\etm{\end{textbmatrix}}

\def\ba#1\ea{\begin{align}#1\end{align}}
\def\bas#1\eas{\begin{align*}#1\end{align*}}
\def\bs#1\es{\begin{split}#1\end{split}}
\def\bg#1\eg{\begin{gather}#1\end{gather}}
\def\bml#1\eml{\begin{multline}#1\end{multline}}
\def\bi#1\ei{\begin{itemize}#1\end{itemize}}

\newcommand{\est}[1]{\ensuremath{\hat{#1}}} 	

\safemath{\dirac}{\delta}					
\safemath{\krond}{\dirac}					

\newcommand{\set}[1]{\left\{#1\right\}}		
\safemath{\upto}{\uparrow}
\safemath{\downto}{\downarrow}
\safemath{\iu}{j}							
\safemath{\ev}{\lambda}						
\safemath{\hilseqspace}{l^{2}}				
\newcommand{\banachfunspace}[1]{\setL^{#1}}	
\safemath{\hilfunspace}{\banachfunspace{2}}	

\safemath{\SNR}{\textit{SNR}} 				
\safemath{\PAR}{\textit{PAR}} 				
\safemath{\No}{N_0}							
\safemath{\Es}{E_s}							
\safemath{\Eb}{E_b}							
\safemath{\EbNo}{\frac{\Eb}{\No}}
\safemath{\EsNo}{\frac{\Es}{\No}}

\DeclareMathOperator{\CHop}{\ensuremath{\opH}} 
\safemath{\tvir}{\rndh_{\CHop}}				
\safemath{\tvtf}{\rndl_{\CHop}}				
\safemath{\spf}{\rnds_{\CHop}}				
\safemath{\bff}{H_{\CHop}}					

\safemath{\ircf}{r_{h}}						
\safemath{\tftvcf}{r_{s}}					
\safemath{\tfcf}{r_{l}}						
\safemath{\bfcf}{r_{H}}						

\safemath{\tcorr}{c_h}						
\safemath{\scf}{c_{s}}						
\safemath{\tfcorr}{c_{l}}					
\safemath{\fcorr}{c_{H}}						

\safemath{\mi}{I}							
\safemath{\capacity}{C}						

\safemath{\normal}{\mathcal{N}}			
\safemath{\jpg}{\mathcal{CN}}			
\safemath{\mchain}{\leftrightarrow}		

\safemath{\dB}{\,\mathrm{dB}}
\safemath{\dBm}{\,\mathrm{dBm}}
\safemath{\Hz}{\,\mathrm{Hz}}
\safemath{\kHz}{\,\mathrm{kHz}}
\safemath{\MHz}{\,\mathrm{MHz}}
\safemath{\GHz}{\,\mathrm{GHz}}
\safemath{\s}{\,\mathrm{s}}
\safemath{\ms}{\,\mathrm{ms}}
\safemath{\mus}{\,\mathrm{\text{\textmu}s}}
\safemath{\ns}{\,\mathrm{ns}}
\safemath{\ps}{\,\mathrm{ps}}
\safemath{\meter}{\,\mathrm{m}}
\safemath{\mm}{\,\mathrm{mm}}
\safemath{\cm}{\,\mathrm{cm}}
\safemath{\m}{\,\mathrm{m}}
\safemath{\W}{\,\mathrm{W}}
\safemath{\mW}{\, \mathrm{mW}}
\safemath{\J}{\,\mathrm{J}}
\safemath{\K}{\,\mathrm{K}}
\safemath{\bit}{\,\mathrm{bit}}
\safemath{\nat}{\,\mathrm{nat}}

\safemath{\define}{\triangleq}			

\safemath{\equivalent}{\sim}
\safemath{\distas}{\sim}					
\safemath{\sdiff}{\Delta}				

\safemath{\reals}{\mathbb{R}}
\safemath{\positivereals}{\reals_{+}}
\safemath{\integers}{\mathbb{Z}}
\safemath{\posint}{\integers_{+}}
\safemath{\naturals}{\mathbb{N}}
\safemath{\posnaturals}{\naturals_{+}}
\safemath{\complexset}{\mathbb{C}}
\safemath{\rationals}{\mathbb{Q}}

\newcommand*{\fancyrefapplabelprefix}{app}		
\newcommand*{\fancyrefthmlabelprefix}{thm}		
\newcommand*{\fancyreflemlabelprefix}{lem}		
\newcommand*{\fancyrefcorlabelprefix}{cor}		
\newcommand*{\fancyrefdeflabelprefix}{def}		
\newcommand*{\fancyrefproplabelprefix}{prop}		
\newcommand*{\fancyrefexmpllabelprefix}{exmpl}
\newcommand*{\fancyrefalglabelprefix}{alg}		
\newcommand*{\fancyreftbllabelprefix}{tbl}		

\frefformat{vario}{\fancyrefseclabelprefix}{Sec.~#1}
\frefformat{vario}{\fancyrefthmlabelprefix}{Thm.~#1}
\frefformat{vario}{\fancyreftbllabelprefix}{Tbl.~#1}
\frefformat{vario}{\fancyreflemlabelprefix}{Lem.~#1}
\frefformat{vario}{\fancyrefcorlabelprefix}{Cor.~#1}
\frefformat{vario}{\fancyrefdeflabelprefix}{Def.~#1}
\frefformat{vario}{\fancyreffiglabelprefix}{Fig.~#1}
\frefformat{vario}{\fancyrefapplabelprefix}{App.~#1}
\frefformat{vario}{\fancyrefeqlabelprefix}{(#1)}
\frefformat{vario}{\fancyrefproplabelprefix}{Prop.~#1}
\frefformat{vario}{\fancyrefexmpllabelprefix}{Ex.~#1}
\frefformat{vario}{\fancyrefalglabelprefix}{Alg.~#1}

%% file: macros/defs.tex
\newcommand{\WiFi}{\mbox{Wi-Fi}\xspace}

\safemath{\dictab}{[\,\dicta\,\,\dictb\,]}

\safemath{\ysig}{\bmy}
\safemath{\ysighat}{\hat{\ysig}}
\safemath{\ysigdim}{M}
\safemath{\xsig}{\bmx}
\safemath{\xsigdim}{N}
\safemath{\nx}{n_x}
\safemath{\zsig}{\bmz}
\safemath{\zsigdim}{\ysigdim}
\safemath{\rsig}{\bmr}
\safemath{\Adict}{\bA}
\safemath{\Adicttilde}{\widetilde{\Adict}}
\safemath{\Adictdim}{\outputdim\times\xsigdim}
\safemath{\avec}{\bma}
\safemath{\avectilde}{\tilde{\avec}}
\safemath{\Bdict}{\bB}
\safemath{\Bdicttilde}{\widetilde{\Bdict}}
\safemath{\Cdict}{\bC}
\safemath{\cvec}{\bmc}
\safemath{\Ddict}{\bD}
\safemath{\Ddictdim}{\ysigdim\times\xsigdim}
\safemath{\dvec}{\bmd}
\safemath{\Ddicttilde}{\widetilde{\bD}}
\safemath{\Bonb}{\bB}
\safemath{\bvec}{\bmb}
\safemath{\Bonbdim}{\ysigdim\times\ysigdim}
\safemath{\noise}{\bmn}
\safemath{\noisedim}{\ysigim}
\safemath{\err}{\bme}
\safemath{\errdim}{\ysigdim}
\safemath{\errset}{\setE}
\safemath{\nerr}{n_e}
\safemath{\delop}{\bP_\errset}
\safemath{\delopc}{\bP_{{\errset}^c}}

\safemath{\cplxi}{\imath}
\safemath{\cplxj}{\jmath}

\safemath{\dict}{\matD}
\safemath{\inputdim}{N}		
\safemath{\outputdim}{M}		
\safemath{\sparsity}{S}	
\safemath{\inputdimA}{{N_a}}	
\safemath{\inputdimB}{{N_b}}	
\safemath{\elemA}{{n_a}}	
\safemath{\elemB}{{n_b}}	
\safemath{\resA}{\matR_a}	
\safemath{\resB}{\matR_b}	
\safemath{\subD}{\matS} 
\safemath{\subA}{\matS_a} 
\safemath{\subB}{\matS_b} 
\safemath{\dicta}{\matA} 	
\safemath{\dictb}{\matB} 	
\safemath{\hollowS}{H}
\safemath{\hollowA}{H_a}
\safemath{\hollowB}{H_b}
\safemath{\cross}{Z}
\safemath{\coh}{\mu_d}			
\safemath{\coha}{\mu_a}			
\safemath{\cohb}{\mu_b}			
\safemath{\mubs}{\nu}	
\safemath{\cohm}{\mu_m} 
\safemath{\dictset}{\setD}	
\safemath{\dictsetp}{\dictset(\coh,\coha,\cohb)}	
\safemath{\dictsetgen}{\dictset_\text{gen}}
\safemath{\dictsetgenp}{\dictsetgen(\coh)}
\safemath{\dictsetonb}{\dictset_\text{onb}}
\safemath{\dictsetonbp}{\dictsetonb(\coh)}

\safemath{\leftside}{U}
\safemath{\rightsideA}{R_a}
\safemath{\rightsideB}{R_b}

\safemath{\indexS}{\setI_S} 

\safemath{\na}{n_a}			
\safemath{\nb}{n_b}			
\safemath{\coeffa}{p_i}	
\safemath{\coeffb}{q_j}	
\safemath{\seta}{\setP}		
\safemath{\setb}{\setQ}     
\safemath{\setw}{\setW}	
\safemath{\setz}{\setZ}	
\safemath{\cola}{\veca}		
\safemath{\colb}{\vecb}		
\safemath{\cold}{\vecd}		
\safemath{\inputvec}{\vecx} 	
\safemath{\error}{\vece}	
\safemath{\noiseout}{\vecz} 	
\safemath{\inputvecel}{x}
\safemath{\inputveca}{\vecx_a}
\safemath{\inputvecb}{\vecx_b}
\safemath{\outputvec}{\vecy}	
\safemath{\lambdamin}{\lambda_{\mathrm{min}}}

\safemath{\elltwo}{\ell_2}
\safemath{\ellone}{\ell_1}
\safemath{\ellzero}{\ell_0}
\safemath{\ellinf}{\ell_\infty}
\safemath{\ellinftilde}{\ell_{\widetilde\infty}}
\safemath{\licard}{Z(\coh,\coha,\cohb)}
\safemath{\xsol}{\hat{x}}
\safemath{\xbord}{x_b}		
\safemath{\xstat}{x_s}		
\safemath{\xstatLone}{\tilde{x}_s}
\safemath{\order}{\mathcal{O}} 
\safemath{\scales}{\Theta} 

\safemath{\ones}{\mathbf{1}} 
\safemath{\zeroes}{\mathbf{0}} 
\safemath{\thlone}{\kappa(\coh,\cohb)} 
\safemath{\constoneA}{\delta} 
\safemath{\constoneB}{\epsilon} 
\safemath{\nlarge}{L}				   
\safemath{\sumlarge}{S_\nlarge}
\safemath{\maxlarger}{P_\nlarge}	   
\safemath{\Pzero}{\textrm{P0}}	
\safemath{\Pone}{\textrm{P1}}
\safemath{\vecfir}{\vecw}			 
\safemath{\vecsec}{\vecz}
\safemath{\elvecfir}{w}              
\safemath{\elvecsec}{z}				 
\safemath{\nlargefir}{n}
\safemath{\normout}{\gamma}
\safemath{\auxfun}{h}
\safemath{\supp}{\textrm{supp}}

\safemath{\indexa}{\ell}
\safemath{\indexb}{r}
\safemath{\indexc}{i}
\safemath{\indexd}{j}

\safemath{\project}{P}